\documentclass{aa}  
\usepackage[version=4]{mhchem}
\begin{document}

   \title{
   Interferometric observations of warm deuterated methanol in the inner regions of low-mass protostars
} 

\titlerunning{Warm methanol deuteration in protostars}

\author{V. Taquet\inst{1} 
\and E. Bianchi\inst{2,1}
\and C. Codella\inst{1,2} 
\and M.~V.~Persson\inst{3} 
\and C. Ceccarelli\inst{2}
\and S. Cabrit\inst{4}
\and J.~K.~J{\o}rgensen\inst{5} 
\and C. Kahane\inst{2}
\and A. L{\'o}pez-Sepulcre\inst{2,6}
\and R. Neri\inst{6}
}

\institute{
INAF, Osservatorio Astrofisico di Arcetri, Largo E. Fermi 5, 50125 Firenze, Italy 
\email{taquet@arcetri.astro.it}
\and IPAG, Université Grenoble Alpes, CNRS, F-38000 Grenoble, France
\and Department of Space, Earth, and Environment, Chalmers University of Technology, Onsala Space Observatory, 439 92 Onsala, Sweden
\and LERMA, Observatoire de Paris, PSL Research University, CNRS, Sorbonne Université, UPMC Univ. Paris 06, 75014 Paris, France
\and Niels Bohr Institute, University of Copenhagen, {\O}ster Voldgade 5–7, DK-1350 Copenhagen K., Denmark
\and Institut de Radioastronomie Millimétrique, 38406 Saint-Martin d’Hères, France
}

   \date{}

 
  \abstract
   {
Methanol is a key species in astrochemistry since it is the most abundant organic molecule in the interstellar medium and is thought to be the mother molecule of many complex organic species. 
Estimating the deuteration of methanol around young protostars is of crucial importance because it highly depends on its formation mechanisms and the physical conditions during its moment of formation.  
We analyse several dozens of transitions from deuterated methanol isotopologues coming from various existing observational datasets obtained with the IRAM-PdBI and ALMA sub-mm interferometers to estimate the methanol deuteration surrounding three low-mass protostars on Solar System scales. A population diagram analysis allows us to derive a [CH$_2$DOH]/[CH$_3$OH] abundance ratio of $3-6$ \% and a [CH$_3$OD]/[CH$_3$OH] ratio of $0.4-1.6$ \% in the warm inner ($\leq 100-200$ AU) protostellar regions. 
These values are typically ten times lower than those derived with previous single-dish observations towards these sources but they are one to two orders of magnitude higher than the methanol deuteration measured in massive hot cores. Dust temperature maps obtained from {\it Herschel} and {\it Planck} observations show that massive hot cores are located in warmer molecular clouds than low-mass sources, with temperature differences of $\sim$10 K. The comparison of our measured values with the predictions of the gas-grain astrochemical model \texttt{GRAINOBLE} shows that such a temperature difference is sufficient to explain the different deuteration observed in low- to high-mass sources. This suggests that the physical conditions of the molecular cloud at the origin of the protostars mostly govern the present observed deuteration of methanol and, therefore, of more complex organic molecules. 
Finally, the methanol deuteration measured towards young solar-type protostars on Solar System scales seems to be higher by a factor of $\sim 5$ than the upper limit in methanol deuteration estimated in comet Hale-Bopp. If this result is confirmed by subsequent observations of other comets, this would imply that an important reprocessing of the organic material likely occurred in the solar nebula during the formation of the Solar System. 
}


   \maketitle
%

\section{Introduction}

Low-mass protostars are known to be chemically rich with the detection of several dozens of molecules \citep{vanDishoeck1995, Cazaux2003, Herbst2009, Caux2011, Caselli2012, Jorgensen2016}. A process of deuterium enrichment is also observed for several neutral species, in particular for molecules thought to be mostly formed in cold interstellar ices and then released in the warm gas surrounding protostars \citep{Ceccarelli2014}. 
So far, methanol and formaldehyde have been the two molecules showing the highest deuterium fractionations, up to 50 \%, or 5 orders of magnitude higher than the cosmic D/H elemental abundance ratio of $1.6 \times 10^{-5}$ \citep{Linsky2003, Ceccarelli1998, Parise2006}. Other species that are thought to be also mostly formed in interstellar ices, such as water, show lower deuteration \citep{Coutens2012, Taquet2013, Persson2014}. These differences have been invoked to reflect different moments or different mechanisms of formation \citep{Cazaux2011, Taquet2013, Taquet2014, Ceccarelli2014}. 

Gas phase chemistry is known to be inefficient for the formation of gaseous methanol \citep{Geppert2006, Garrod2006}. Instead, methanol has been found to be efficiently formed at the surface of cold interstellar grains in dark cloud conditions through the hydrogenation of CO and H$_2$CO \citep{Watanabe2002, Rimola2014}. 
During the collapse of the protostellar envelope, solid methanol is released into the gas phase once temperatures exceed $\sim 100$ K. 
The deuteration of methanol is thought to remain constant after its evaporation in the warm gas phase of the inner protostellar regions because gas phase chemistry is likely too slow to significantly alter the ratios during the lifetime of the protostar \citep{Charnley1997, Osamura2004}. As a consequence, the deuteration observed around low-mass protostars would be a fossil of its formation in the precursor dense cloud where interstellar ices formed. 
\citet{Taquet2012, Taquet2013} showed that, as other species, methanol deuteration is highly sensitive to the temperature and the density during ice formation. A deuteration higher than 1 \% is favoured by a methanol formation in cold ($T=10$ K) and dense ($n_{\rm H} \geq 10^5$ cm$^{-3}$) molecular cloud conditions. 

Most measurements of the methanol deuteration carried out so far have been performed using single-dish millimeter facilities with beams larger than 10$\arcsec$ (or $\sim 2000$ AU at a typical distance of 200 pc). Such large beams encompass the different components of the protostellar system: the small hot-core where all the icy content is thermally evaporated but also the cold external envelope and molecular outflows driven by the central source(s). 
The advent of modern millimeter interferometers, such as the IRAM Plateau de Bure Interferometer (IRAM-PdBI) now called the NOrthern Extended Millimeter Array (NOEMA) and the Atacama Large Millimeter/submillimeter Array (ALMA) now allows astronomers to zoom towards the inner protostellar regions, in
the so-called hot corinos, to measure the deuteration of the whole icy content released in the gas phase.
New measurements towards low-mass protostars within different nearby molecular clouds suggest lower methanol deuterations than those originally measured a decade ago with single-dish telescopes.
\citet{Bianchi2017} derived a [CH$_2$DOH]/[CH$_3$OH] abundance ratio of 2 \% in HH212, a low-mass protostar located in the Orion molecular cloud, whilst \citet{Jorgensen2018} estimated a [CH$_2$DOH]/[CH$_3$OH] ratio of 7 \% in Source B of IRAS 16293-2422, a low-mass protobinary system located in the Ophiuchus cloud.
In addition to being lower with respect to the previous single-dish measurements of $20-60$ \%, they also differ within each other by a factor of $\sim 3$. Such a difference could be due to different properties of the precursor clouds. The Orion molecular cloud is the closest massive star-forming region and is known to be more active and slightly warmer than the more quiescent Ophiuchus cloud. However, this assumption remains to be confirmed for sources located in other clouds. 

This article presents an analysis of different published observational datasets from the IRAM-PdBI towards NGC1333-IRAS2A and -IRAS4A, two bright Class 0 low-mass protostars located in the nearby Perseus molecular cloud. We also re-analyse partially published data from the ALMA towards HH212 in order to measure the [CH$_2$DOH]/[CH$_3$OH] and [CH$_3$OD]/[CH$_3$OH] abundance ratios in their hot corinos. 
{In this article, we aim to investigate whether the warm methanol deuteration observed in hot cores is regulated by the dust temperature of precursor dark clouds. To this aim, we compare our data with interferometric observations towards more massive protostellar systems and with the predictions of a state-of-the-art astrochemical model. }

\section{Observations}

\subsection{Observational details}

The two low-mass Class 0 protostars IRAS2A and IRAS4A located in the NGC1333 cloud {at $299 \pm 14$ pc \citep{Zucker2018}}
were observed with the IRAM PdBI at 143, 165, and 225 GHz.
Observations at 143 and 165 GHz were carried out on July 20, July
21, August 1, August 3, November 24 2010 and March 10 2011 in the C and D
configurations of the array. 
Observations at 225 GHz were performed on 27 and 28 November 2011,
on 12, 15, 21, 27 March 2012, and on 2 April 2012 in the B and C configurations of the array.   
A more detailed description of the observational setups is provided in \citet{Taquet2015} and \citet{Persson2014}, respectively.
The amplitude calibration uncertainty is estimated to be $\sim$20\%. 
The WIDEX correlator has been used at three frequency settings,
providing a bandwidth of 3.6 GHz each with a spectral resolution
of 1.95 MHz (4, 3.5, and 2.6 km s$^{-1}$ at 145, 165, and 225 GHz, respectively). 
The data calibration and imaging were performed using the CLIC and
MAPPING packages of the GILDAS software
\footnote{http://www.iram.fr/IRAMFR/GILDAS}. 
Continuum images were produced by averaging line-free channels in the
WIDEX correlator before the Fourier transformation of the data.
The line spectral cubes were then obtained by subtracting the
continuum visibilities from the whole (line+continuum) datacube,
followed by natural weighted cleaning of individual channels. 
For IRAS4A, the baseline has also been flattened by importing the data
cubes into CLASS and subtracting a polynomial function of low order to each
individual spectrum.
{Positions are given with respect to the continuum peaks of IRAS2A and IRAS4A1 located respectively at
$\alpha({\rm J2000})$ = 03$^h$ 28$^m$ 55$\fs$57, $\delta({\rm J2000})$ = +31$\degr$ 14$\arcmin$ 37$\farcs$22 and 
$\alpha({\rm J2000})$ = 03$^h$ 29$^m$ 10$\fs$52, $\delta({\rm J2000})$ = +31$\degr$ 13$\arcmin$ 31$\farcs$06.
}

The HH212-MM1 (hereafter HH212) protostar located in the Orion B cloud at about 
400 pc \citep{Kounkel2017} was observed in Band 7 with ALMA using 
34 12-m antennas between June 15 and July 19 2014 during the Cycle 1 phase and 
44 12-m antennas between October 6 and November 26 2016 during the Cycle 4 phase.
The maximum baselines for the Cycle 1 and 4 observations are 650 m and 3 km, respectively.
In Cycle 1, the spectral windows between $337.1-338.9$ GHz and $348.4-350.7$ GHz 
were observed using spectral channels of 977 kHz (or $0.87$ km s$^{-1}$), 
subsequently smoothed to $1.0$ km s$^{-1}$ to increase sensitivity. 
Calibration was carried out following standard procedures, using quasars J0607-0834, 
J0541–0541, J0423–013, and Ganymede.
For the Cycle 4 data, a single spectral window between 334.1--336.0 GHz with 
a 488 kHz (or $0.42$ km s$^{-1}$) spectral resolution, successively smoothed to
1 km s$^{-1}$ was used.
Calibration was carried out using quasars J0510+1800, J0552+0313, J0541--0211 and J0552--3627.
Spectral line imaging was achieved using CASA\footnote{http://casa.nrao.edu}, 
whilst data analysis was performed using the GILDAS package.
Positions are given with respect to the MM1 protostar continuum peak
located at $\alpha({\rm J2000})$ = 05$^h$ 43$^m$ 51$\fs$41,
$\delta({\rm J2000})$ = --01$\degr$ 02$\arcmin$ 53$\farcs$17.
In addition to $^{13}$CH$_{\rm 3}$OH and CH$_{\rm 2}$DOH images already presented by \citet{Bianchi2017}, 
we also report the imaging of the CH$_{\rm 3}$OD(6$_{\rm 2-}$--6$_{\rm 1+}$) transition at 335.1 GHz \citep[see also][]{Codella2019}.

The size of the synthesized beams and the rms noise are reported in
Table \ref{prop_cont}.  

\begin{table}[htp]
\centering
\caption{Parameters of the observational data.}
\begin{tabular}{l c c c}
\hline
\hline
 & \multicolumn{3}{c}{NGC1333-IRAS2A (PdBI)} \\
\hline
Frequency & 145 GHz & 165 GHz & 225 GHz \\
 \hline
Beam size (\arcsec) & 2.1$\times$1.7 & 2.3$\times$1.7 & 1.2$\times$1.0 \\
Beam PA ($\degr$) & 25 & 110 & 22 \\
rms(spectrum) $^a$ & 2.6 & 3.5 & 2.9 \\
\hline
 & \multicolumn{3}{c}{NGC1333-IRAS4A (PdBI)} \\
\hline
Frequency & 145 GHz & 165 GHz & 225 GHz \\
 \hline
Beam size (\arcsec) & 2.2$\times$1.7 & 2.4$\times$1.8 & 1.1$\times$0.8 \\
Beam PA ($\degr$) & 25 & 114 & 14 \\
rms(spectrum) $^a$ & 3.3 & 4.0 & 4.8 \\
\hline
 & \multicolumn{3}{c}{HH212 (ALMA)} \\
\hline
 & 335 GHz & 338 GHz & 349 GHz \\
 \hline
Beam size (\arcsec) & 0.15$\times$0.12 & 0.52$\times$0.34 & 0.41$\times$0.33 \\
Beam PA ($\degr$) & -88 & -63 & -63 \\
rms(spectrum) $^b$ & 1 & $5-6$ & $5-6$ \\
\hline
\end{tabular}
\tablebib{
$a$: Units of mJy beam$^{-1}$ channel$^{-1}$ for a channel width of 1.95 MHz. 
$b$: Units of mJy beam$^{-1}$ channel$^{-1}$ for a channel width of 0.87 km s$^{-1}$ (335 GHz) and 1.0 km s$^{-1}$ (at 338 and 349 GHz). 
}
\label{prop_cont}
\end{table}

\subsection{Spectroscopic parameters} \label{spec_params}

In this work, we focus on the CH$_2$DOH, CH$_3$OD, and CHD$_2$OH deuterated isotopologues. 
We use the CH$_2$DOH JPL data entry from \citet{Pearson2012}. The new band strengths published by \citet{Pearson2012} can differ by a factor of a 1.5 - 2.0 with respect to the values from \citet{Parise2002} used in most previous studies before 2012. With a vibrational correction factor of 1.15 at 160 K and 1.46 at 300 K based on the torsional data of \citet{Lauvergnat2009}, the partition function is 1.3 - 2.0 higher than the partition function estimated by \citet{Parise2002}. 
We re-analyse the CH$_2$DOH transitions observed towards the IRAS 16293-2422 low-mass protostellar system with the IRAM 30 meters telescope by \citet{Parise2002} to investigate the impact of these new spectroscopic data on the estimation of the CH$_2$DOH column densities. We derive a CH$_2$DOH column density 2.2 lower than the one found by these authors. 
For CH$_3$OD, we prepare a catalog entry based on the band strengths published by \citet{Anderson1988} with frequencies updated by \citet{Duan2003}. 
{A self-consistent calculation of the CH$_3$OD partition function has not be performed so far since the CH$_3$OD spectrum has only been partially studied in the laboratory. Instead,}
we use two estimates for the CH$_3$OD partition function $Z$ to derive the CH$_3$OD column densities. The first one follows the estimation by \citet{Parise2004} within the rigid-rotor approximation, giving $Z = 1.42 T^{3/2}$ \citep[see also][]{Ratajczak2011} used in most previous studies. The second one follows the estimation by \citet{Belloche2016} whose partition function values are scaled with those of the CH$_3^{18}$OH partition function \citep[see also][]{Jorgensen2018}. {At 150 K, the the partition function estimated by \citet{Parise2004} is four times lower than the one estimated by \citet{Jorgensen2018}}, resulting in large differences in the derived CH$_3$OD column densities. 
For CHD$_2$OH, we use the spectroscopic data and the partition function published by \citet{Parise2002}. The CHD$_2$OH partition function is found to be similar to that of CH$_2$DOH published by the JPL database with differences lower than 20 \% between 10 and 300 K. 

We also target transitions from CH$_2$DCN. The CH$_2$DCN spectroscopic data are taken from the CDMS catalog, based on \citet{Nguyen2013}. 

\begin{figure*}[htp]
\centering
\includegraphics[width=150mm]{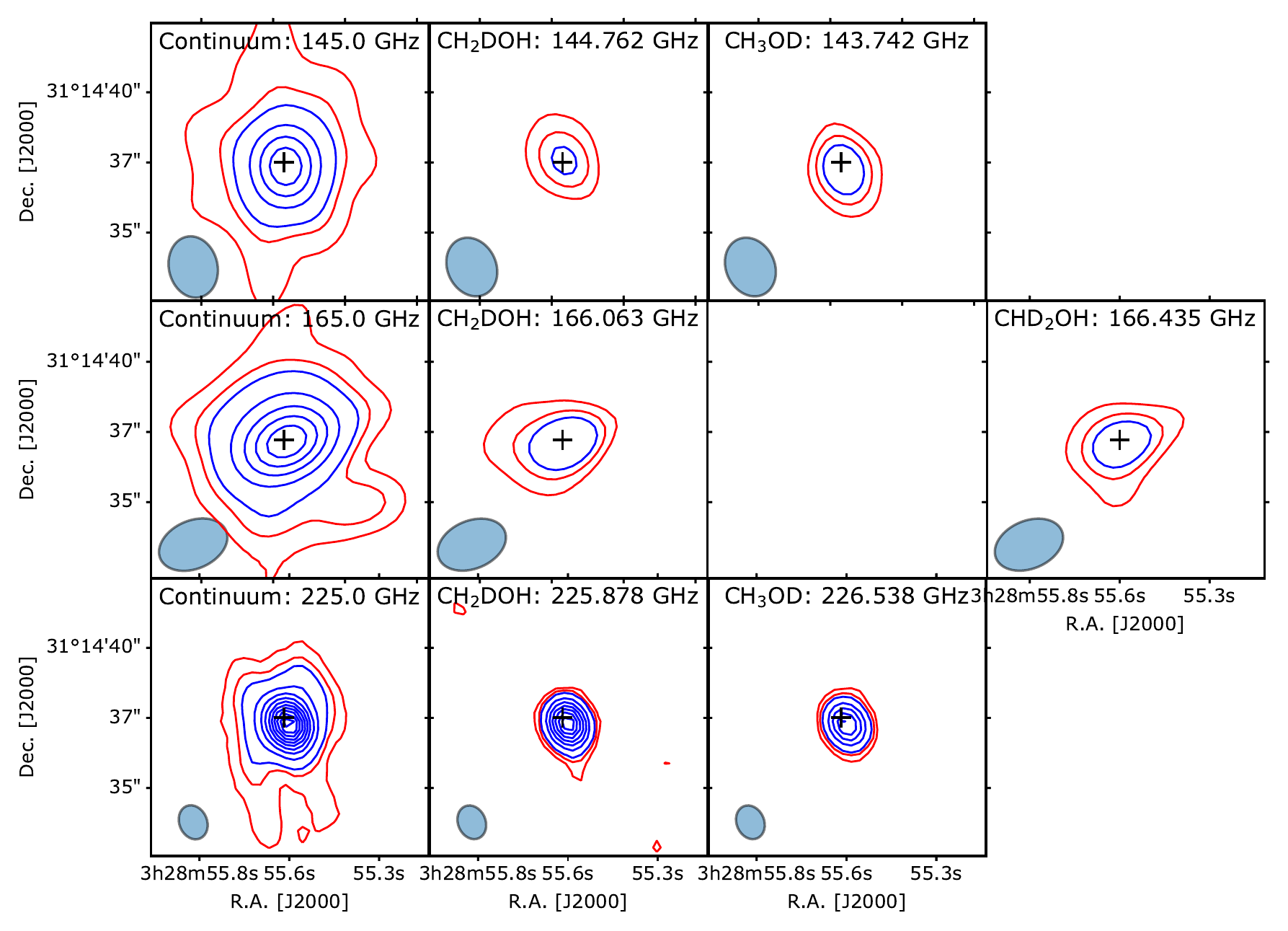} 
\caption{Integrated maps of the continuum, CH$_2$DOH, CH$_3$OD, and CHD$_2$OH emission observed towards
  IRAS2A for the three PdBI frequency settings (top: 145 GHz, middle: 165 GHz, bottom: 225 GHz). Red contours show the the 3$\sigma$ and
  6$\sigma$ levels, whilst blue contours are in step of 9$\sigma$. 
The synthesized beams are shown in the bottom left of each panel. The black cross depicts the position of the protostar.}
\label{maps_iras2a}
\end{figure*}

\begin{figure*}[htp]
\centering
\includegraphics[width=150mm]{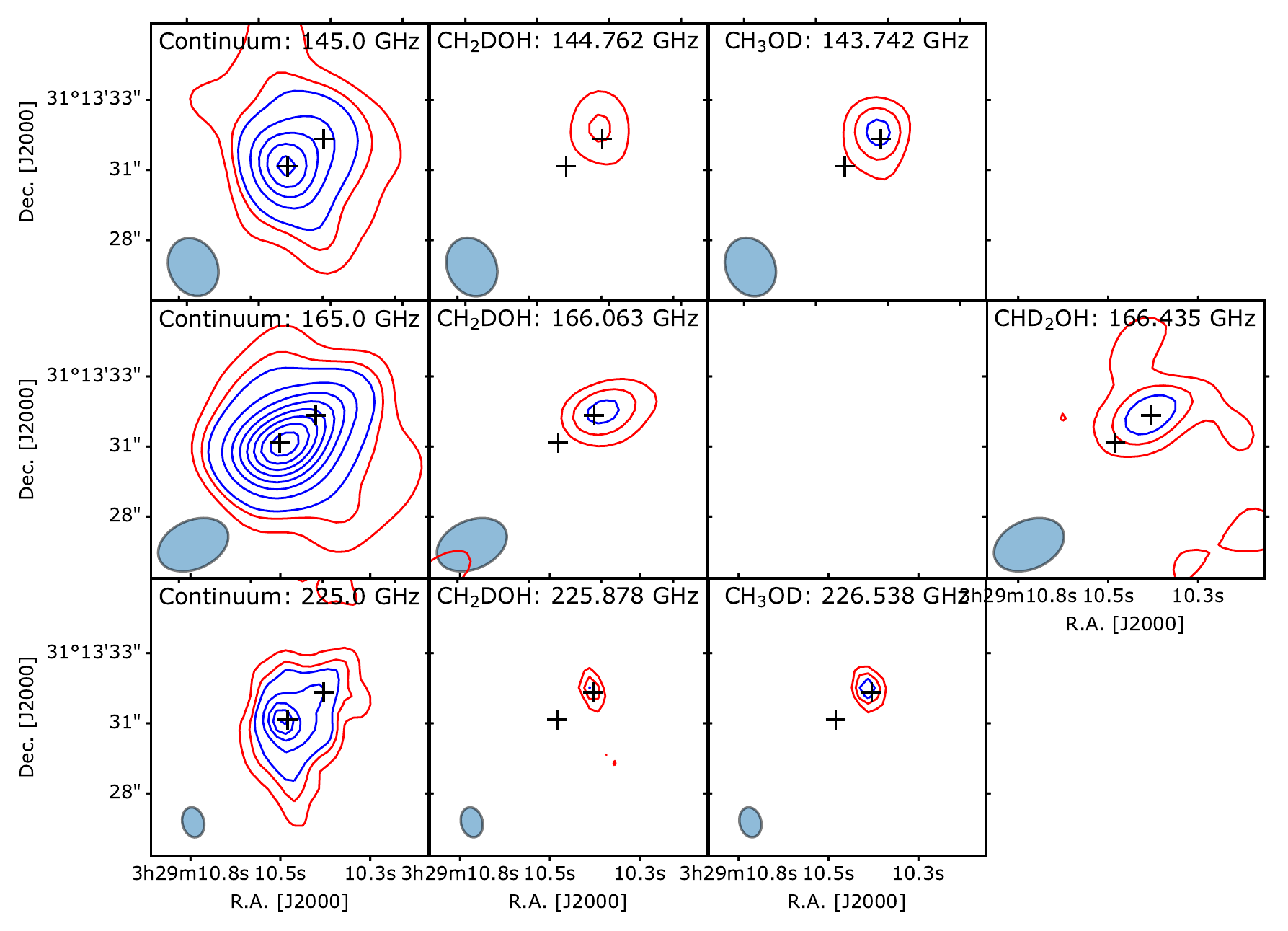} 
\caption{Integrated maps of the continuum, CH$_2$DOH, CH$_3$OD, and CHD$_2$OH emission observed towards
  IRAS4A for the three PdBI frequency settings (top: 145 GHz, middle: 165 GHz, bottom: 225 GHz). Red contours show the the 3$\sigma$ and
  6$\sigma$ levels, whilst blue contours are in step of 9$\sigma$. 
The synthesized beams are shown in the bottom left of each panel. The black crosses depict the position of the protostars.}
\label{maps_iras4a}
\end{figure*}

\begin{figure*}[htp]
\centering
\includegraphics[width=117.0mm]{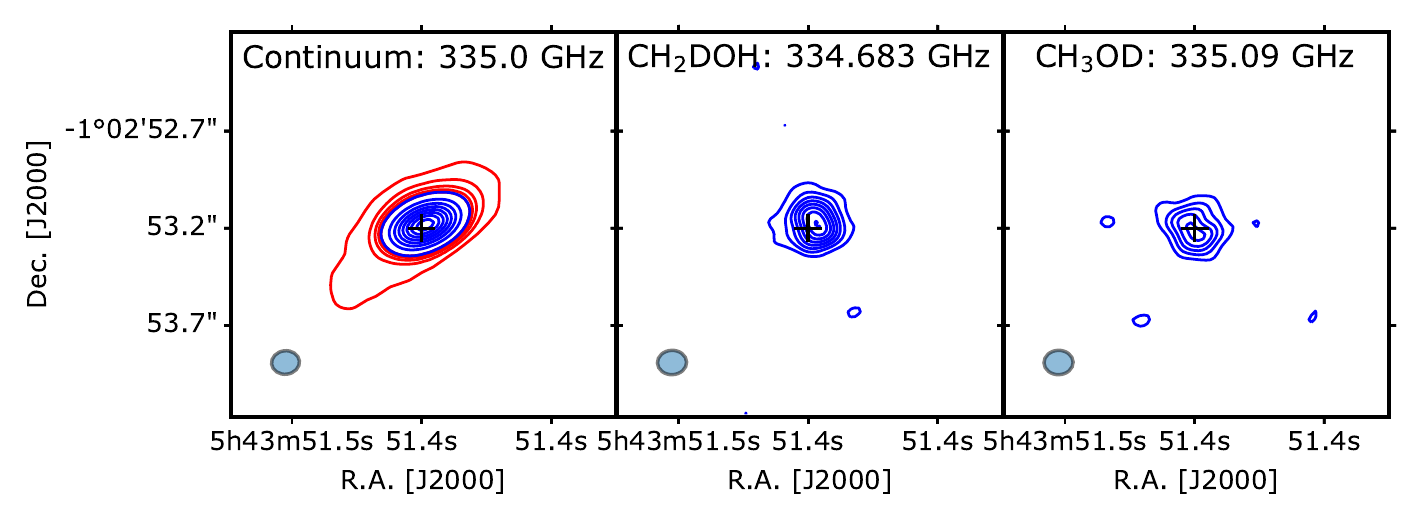} 
\caption{Integrated maps of the continuum , CH$_2$DOH, and CH$_3$OD emission observed towards
  HH212 for the ALMA frequency setting at 335 GHz. 
  For the continuum map, red contours show the 5 first 20$\sigma$ levels and blue contours are in steps of 100$\sigma$.
  For the molecular maps, blue contours are in steps of 3$\sigma$.
The synthesized beams are shown in the bottom left of each panel. }
\label{maps_hh212}
\end{figure*}

\section{Results}

\subsection{Integrated maps}

For IRAS2A and IRAS4A, the integrated emission maps of all methanol isotopologue transitions are obtained by integrating the flux over $V_{\rm sys} \pm \Delta V$, where $V_{\rm sys}$ is the source systemic velocity and $\Delta V$ is the transition FWHM linewidth. In practice, we integrate the line emission over three channels given the low spectral resolution of the WIDEX data. 
Figures \ref{maps_iras2a} and \ref{maps_iras4a} give an overview of the maps obtained for CH$_2$DOH, CH$_3$OD, and CHD$_2$OH in the three frequency settings. 
For all isotopologues, the emission is limited to the inner regions around the protostars within the synthesized beams. Towards IRAS4A, the molecular emission originates from the north-western source IRAS4A2 although the continuum emission of the sourth-eastern source IRAS4A1 is brighter. These maps are in good agreement with previous interferometric observations of these sources that suggest that the emission of complex organic molecules originates from "hot corinos" whose sizes are smaller than 0$\farcs$5 \citep[or 120 au;][]{Jorgensen2005, Persson2012, Maury2014, Taquet2015, Sepulcre2017, DeSimone2017}. 

For HH212, we produce the integrated emission maps of $^{13}$CH$_3$OH, CH$_2$DOH, and CH$_3$OD transitions that are not blended with other transitions by integrating the flux over the entire linewidth after an analysis of their spectra towards the continuum peak. 
The Cycle 1 emission is spatially unresolved with a a beam of 0\farcs5$\times$0\farcs3, corresponding to a size of 225$\times$135 au. However, the analysis of the higher angular resolution Cycle 4 dataset provides a beam-deconvolved FWHM size of 0\farcs18$\times$0\farcs12 (81$\times$54 au) for the $^{13}$CH$_3$OH emission and 0\farcs20$\times$0\farcs10 (90$\times$45 au) for CH$_2$DOH. The emission originates from the “hot-corino” region and is confined in a rotating structure that extends at  $\pm 45$ au from the equatorial plane and is elongated along and rotates around the jet axis. The actual interpretation of this emission is still under debate and has been possibly attributed to disc wind, or disc atmosphere from accretion shocks \citep{Bianchi2017, Lee2017}.

We measure the flux of all methanol isotopologue transitions integrated over an elliptical mask with similar size and position angle to the synthesized beam and with a center at the molecular emission peak following the methodology described in \citet{Taquet2015}. Tables \ref{lines_ch2doh_2a4a} to \ref{lines_ch3od_hh212} of the Appendix summarise the properties and the measured flux of the targeted transitions towards the two sources.

\subsection{Population diagrams}

\begin{figure}[htp]
\centering
\includegraphics[width=\columnwidth]{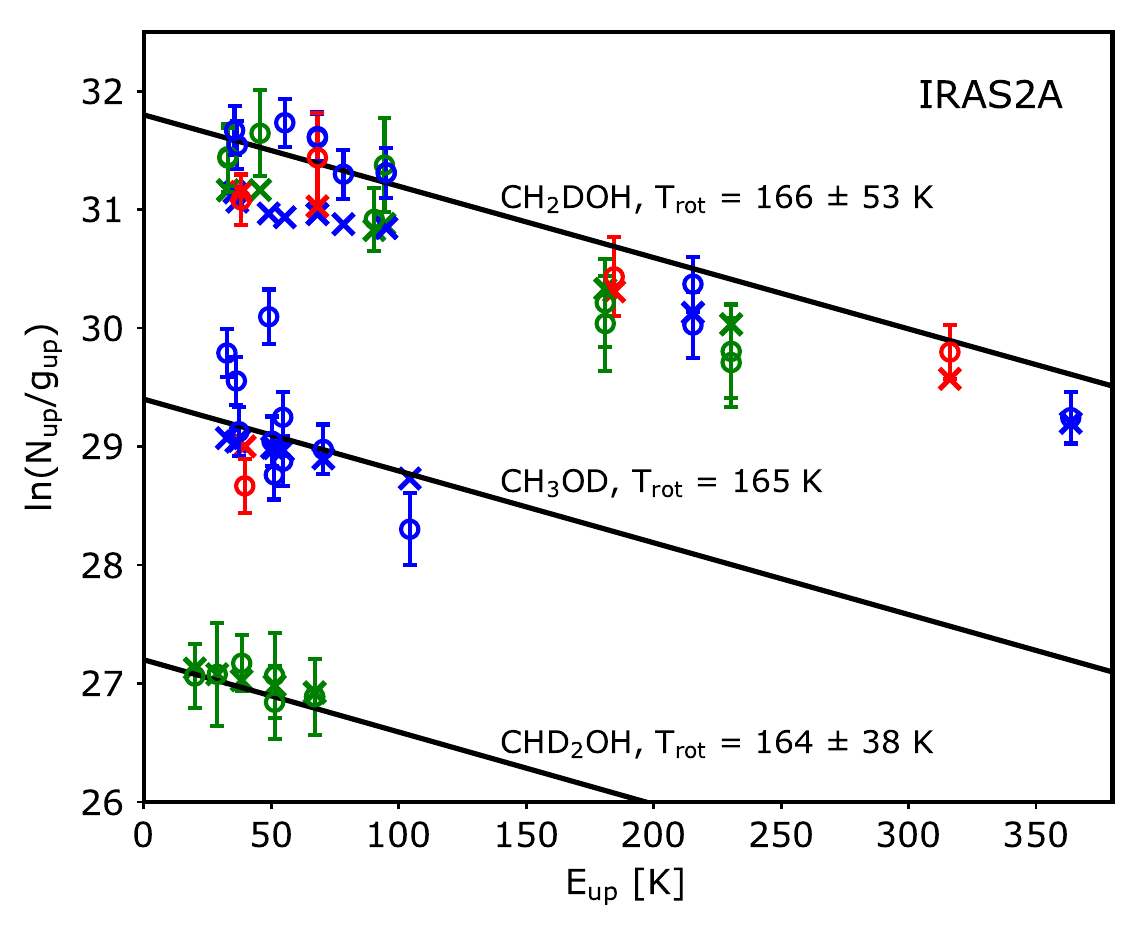} 
\caption{Rotational and population diagrams of deuterated methanol isotopologues
  (CH$_2$DOH, CH$_3$OD, and CHD$_2$OH) derived towards NGC1333-IRAS2A for 
  source sizes derived from the PD analysis of the methanol population
  distribution \citep[0$\farcs$36; see][]{Taquet2015}. 
  Red, green, and blue symbols represent transitions at 145, 165, and 225 GHz, 
  respectively.
  The CH$_3$OD and CHD$_2$OH levels have been artificially shifted by -1.5 and -4 for sake of clarity.
  Observational data are depicted by the diamonds. Crosses show the
  best fit of the PD to the data. Blue and green symbols represent
  transitions at 2 and 1.3 mm, respectively. Error bars are derived assuming a calibration
  uncertainty of 20 \% on top of the statistical error.
  Black straight lines represent the best fit of the rotational diagram analysis, assuming optically thin emission, to the data. 
  }
\label{PD_iras2a}
\end{figure}

\begin{figure}[htp]
\centering
\includegraphics[width=\columnwidth]{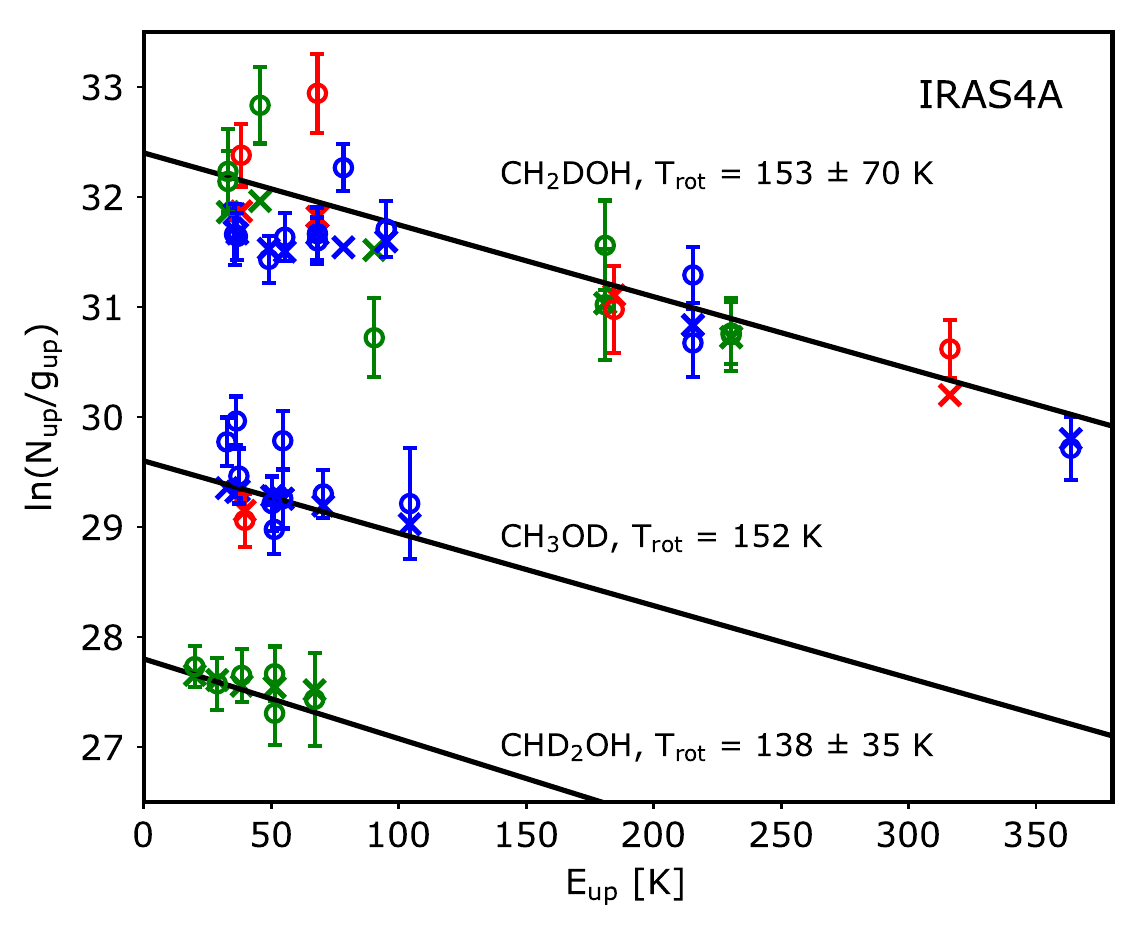} 
\caption{
  Rotational and population diagrams of deuterated methanol isotopologues
  (CH$_2$DOH, CH$_3$OD, and CHD$_2$OH) derived towards NGC1333-IRAS2A for 
  source sizes derived from the PD analysis of the methanol population
  distribution \citep[0$\farcs$20; see][]{Taquet2015}. 
  Red, green, and blue symbols represent transitions at 145, 165, and 225 GHz, 
  respectively.
  The CH$_3$OD and CHD$_2$OH levels have been artificially shifted by -1.5 and -4 for sake of clarity.
  Observational data are depicted by the diamonds. Crosses show the
  best fit of the PD to the data. Blue and green symbols represent
  transitions at 2 and 1.3 mm, respectively. Error bars are derived assuming a calibration
  uncertainty of 20 \% on top of the statistical error.
  Black straight lines represent the best fit of the rotational diagram analysis, assuming optically thin emission, to the data. 
  }
\label{PD_iras4a}
\end{figure}

\begin{figure}[htp]
\centering
\includegraphics[width=\columnwidth]{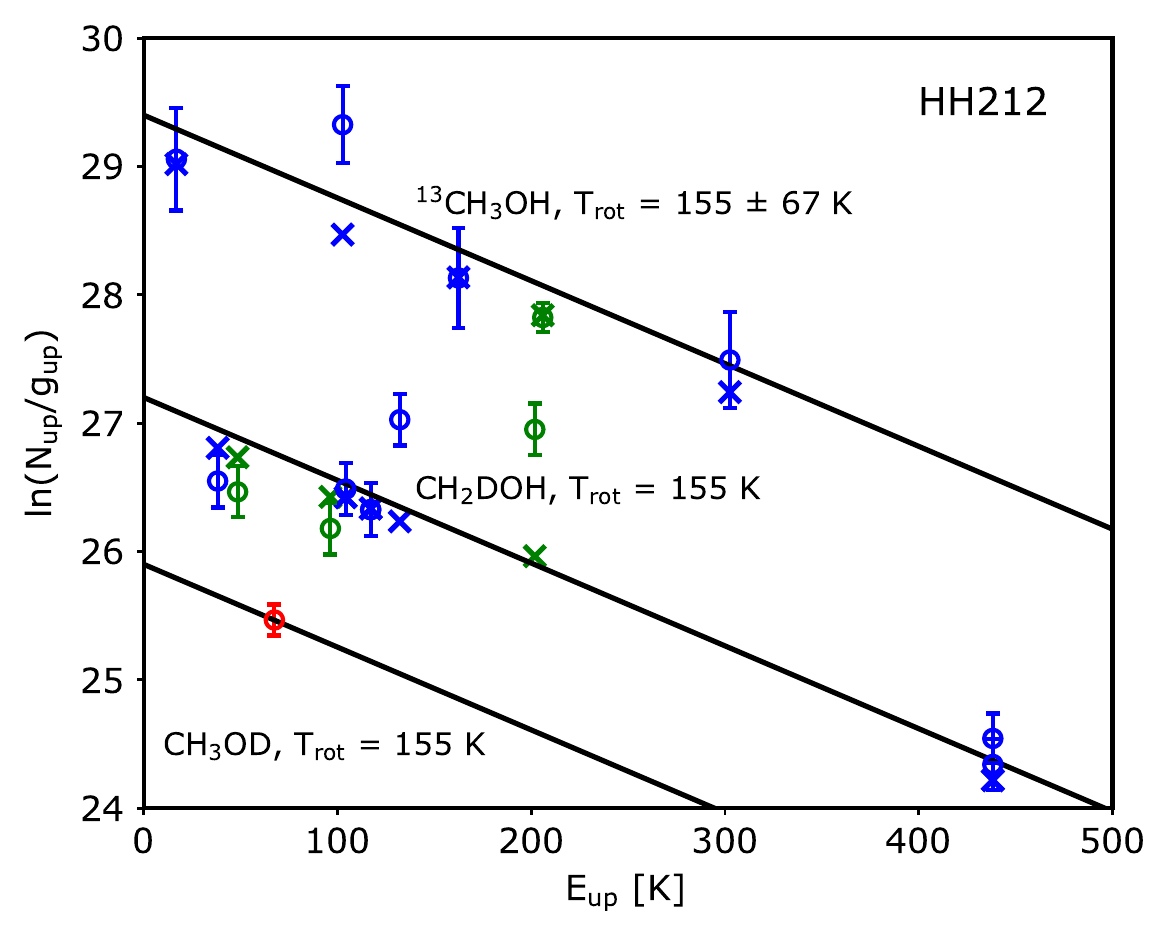} 
\caption{Rotational and population diagrams of methanol isotopologues
  ($^{13}$CH$_3$OH, CH$_2$DOH, and CH$_3$OD) derived towards HH212 for 
  the source sizes derived from the ALMA Cycle 4 data \citep[0$\farcs$19; see][]{Bianchi2017}. 
  The CH$_2$DOH and CH$_3$OD levels have been artificially shifted by -3 for sake of clarity.
  Red, green, and blue symbols represent transitions at 335, 338, and 349 GHz, respectively.
  Observational data are depicted by the diamonds. Crosses show the best fit of the PD to the data. 
  Black straight lines represent the best fit of the rotational diagram analysis, assuming optically thin emission, to the data. 
  }
\label{PD_hh212}
\end{figure}

\begin{table*}[htp]
\centering
\caption{Results from the population diagram analysis for the methanol
  and methyl cyanide deuterated isotopologues.}
\begin{footnotesize}
\begin{tabular}{l c c c c c}
\hline                                    
\hline                                                          
Molecule  &   $N_{hc}$            &   $T_{rot}$ &   Source size         &   $N$(XD)/$N$(XH) $^a$    & $N$(XD)/$N$(XH) $^b$ \\
  &   (cm$^{-2}$) &           (K)         &   (\arcsec)         &   (\%)        & (Single-dish, \%) \\
\hline                                                      
\multicolumn{6}{c}{IRAS2A}  \\
\hline                                                      
CH$_3$OH        & ($5.0_{-1.8}^{+2.9}$)(+18)  & $140_{-20}^{+20}$    & $0.36_{-0.04}^{+0.04}$  $^d$  &    -     & -  \\  
$^{13}$CH$_3$OH & ($7.1_{-2.6}^{+4.2}$)(+16)  & $140_{-20}^{+20}$    & $0.36_{-0.04}^{+0.04}$  $^d$  &    -     &  - \\ 
CH$_2$DOH       & ($2.9_{-0.6}^{+1.2}$)(+17)  & $166_{-48}^{+58}$    & 0.36          $^e$            & $5.8_{-1.5}^{+1.3}$ & $62_{-33}^{+71}$ \\  
CH$_3$OD (P04$^c$)  & ($7.9_{-1.6}^{+2.1}$)(+16)  & 166         $^f$     & 0.36          $^e$            & $1.6_{-0.8}^{+0.6}$ & $\leq 8$ \\  
CH$_3$OD (J18$^c$)  & ($3.6_{-0.4}^{+0.9}$)(+17)  & 166         $^f$     & 0.36          $^e$            & $7.1_{-1.6}^{+1.2}$ & $\leq 8$ \\ 
CHD$_2$OH       & ($2.2_{-0.5}^{+0.6}$)(+17)  & $164_{-32}^{+44}$    & 0.36          $^e$            & $4.4_{-1.3}^{+0.9}$ & $25_{-14}^{+29}$  \\ 
 \hline
CH$_3$CN        & ($2.0_{-0.4}^{+1.2}$)(+16)  & $130_{-40}^{+230}$   & 0.36          $^d$            &    -     &  - \\ 
CH$_2$DCN       & ($7.2_{-0.9}^{+1.5}$)(+14)  & 130         $^g$     & 0.36          $^g$            & $3.6_{-1.2}^{+0.5}$ & - \\ 
\hline                                                        
\multicolumn{6}{c}{IRAS4A}  \\  
\hline                                                        
CH$_3$OH        & ($1.6_{-0.8}^{+0.6}$)(+19) & $140_{-30}^{+30}$    & $0.20_{-0.04}^{+0.08}$  $^d$  &   -     &  -  \\  
$^{13}$CH$_3$OH & ($2.3_{-1.1}^{+1.3}$)(+17) & $140_{-30}^{+30}$    & $0.20_{-0.04}^{+0.08}$  $^d$  &   -     &  -  \\  
CH$_2$DOH       & ($5.9_{-1.8}^{+2.2}$)(+17) & $152_{-78}^{+62}$    & 0.20          $^e$          & $3.7_{-0.9}^{+1.2}$  & $65_{-21}^{+30}$ \\  
CH$_3$OD (P04$^c$)  & ($1.1_{-0.2}^{+0.5}$)(+17) & 152         $^f$     & 0.20          $^e$          & $0.7_{-0.4}^{+0.5}$  & $4.7_{-2.1}^{+2.9}$ \\  
CH$_3$OD (J18$^c$)  & ($5.0_{-1.0}^{+1.3}$)(+17) & 152         $^f$     & 0.20          $^e$          & $3.1_{-0.8}^{+1.0}$  & $4.7_{-2.1}^{+2.9}$ \\  
CHD$_2$OH      & ($3.3_{-0.7}^{+0.9}$)(+17)  & $138_{-38}^{+34}$    & 0.20          $^e$          & $2.1_{-0.6}^{+0.8}$ & $17_{-6}^{+9}$ \\  
 \hline
CH$_3$CN  & ($6.3_{-1.3}^{+3.6}$)(+16)       & $200_{-40}^{+110}$   & 0.20          $^d$          &   -     &  - \\
CH$_2$DCN & ($1.7_{-0.7}^{+0.7}$)(+15)      & 200         $^g$     & 0.20          $^g$          & $2.7_{-1.1}^{+0.7}$ & - \\
\hline                                                        
\multicolumn{6}{c}{HH212}  \\  
\hline                                                        
$^{13}$CH$_3$OH & ($3.2_{-1.4}^{+1.4}$)(+16) & $155_{-65}^{+69}$    & 0.19  $^h$  &   -      &  - \\  
CH$_2$DOH       & ($6.4_{-1.3}^{+1.3}$)(+16) & 155 $^i$    & 0.19  $^h$         & $2.9_{-0.8}^{+0.8}$ & -\\  
CH$_3$OD (P04$^c$)  & ($9.9_{-1.2}^{+1.2}$)(+15) & 155 $^i$    & 0.19  $^h$         & $0.4_{-0.3}^{+0.3}$ &- \\  
CH$_3$OD (J18$^c$)  & ($4.4_{-0.5}^{+0.5}$)(+16) & 155 $^i$    & 0.19  $^h$         & $2.0_{-0.6}^{+0.6}$ & -\\  
\hline                                  
\end{tabular}
\label{recap_PD}
\tablebib{
{\bf a}: Column density ratio between a deuterated species and its main isotopologue. 
{\bf b}: From \citet{Parise2006}. 
{\bf c}: P04 refers to the partition function computed by \citet{Parise2004} within the rigid-rotor approximation whilst J18 refers to the partition function used in \citet{Jorgensen2018} obtained from a scaling of that of CH$_3^{18}$OH. 
{\bf d}: From \citet{Taquet2015}. \\
{\bf e}: The source size was assumed to be equal to that of CH$_3$OH. 
{\bf f}: The rotational temperature was assumed to be equal to that of CH$_2$DOH. 
{\bf g}: The rotational temperature was assumed to be equal to that of CH$_3$CN. 
{\bf h}: From \citet{Bianchi2017}. 
{\bf i}: The rotational temperature was assumed to be equal to that of $^{13}$CH$_3$OH. 
}
\end{footnotesize}
\end{table*}

We use the so-called Population Diagram (PD) method described in \citet{Goldsmith1999} and \citet{Taquet2015} to investigate the
effect of optical depth on the column densities of each level for the different methanol isotopologues. 
We perform a reduced chi-square ($\chi_{\textrm{red}}^2$) minimization by running a grid of models covering a large parameter space in rotational temperature $T_{\textrm{rot}}$ between 50 and 350 K, and total column density in the source $N_{\textrm{tot}}$ between $10^{15}$ and $10^{20}$ cm$^{-2}$. We assume the source sizes $\theta_s$ derived in \citet{Taquet2015} and \citet{Bianchi2017}.
A source size of 0$\farcs$36 and 0$\farcs$20 in IRAS2A and IRAS4A, respectively, was needed to reproduce the scatter of the population distribution of the low CH$_3$OH upper energy levels \citep[see][]{Taquet2015} and 
a source size of 0$\farcs$19 was derived by \citet{Bianchi2017} from the analysis of the spatially resolved $^{13}$CH$_3$OH Cycle 4 emission map.  
At LTE, the column density of every upper state $N_{\textrm{up}}$ can be derived for each set of $N_{\textrm{tot}}$, $T_{\textrm{rot}}$, and source solid angle $\Omega_s$. 
The best-fit model populations are plotted together with the observed populations of the levels in Figures \ref{PD_iras2a}, \ref{PD_iras4a}, and \ref{PD_hh212} and are marked by cross symbols. Table \ref{recap_PD} summarises the parameters of the best-fit models and their associated uncertainties. 
{The column densities of CH$_3$OH and CH$_3$CN have been derived in \citet{Taquet2015} via the analysis of the emission from 28 CH$_3$OH and 13 $^{13}$CH$_3$OH transitions assuming a $^{12}$C/$^{13}$C elemental ratio of 70 and from six CH$_3$CN transitions detected in the 145 and 165 GHz frequency settings, respectively. \citet{Taquet2015} used the same Population Diagram method by considering the rotational temperature $T_{\textrm{rot}}$, the total column density in the source $N_{\textrm{tot}}$, and the source size $\theta_{\rm S}$ as free parameters. }

For IRAS2A and IRAS4A, we start the analysis with CH$_2$DOH since this is the isotopologue with the highest number of detected transitions and with the largest range of excitation, with 25 detected transitions of upper level energies $E_{\rm up}$ between 33 and 364 K. 
The population distributions can be reproduced with rotational temperatures of $166_{-48}^{+58}$ and $152_{-78}^{+62}$ K in IRAS2A and IRAS4A, respectively. These temperatures are slightly higher, but within the uncertainties, than the temperature of 140 K needed to reproduce the CH$_3$OH and $^{13}$CH$_3$OH emissions \citep{Taquet2015}. 
Transitions with upper level energies lower than 100 K are optically thick with opacities of 0.2 - 0.4 towards IRAS2A and 0.5 - 1.0 towards IRAS4A, depending on their properties. 
For CH$_3$OD, we assume the same rotational temperature than CH$_2$DOH because of the lower range of excitation ($E_{\rm up}$ from 33 to 104 K) and the high scatter and uncertainties in the population distributions for both sources. 
The strong scatter of the CH$_3$OD population is not entirely reproduced, possibly because of non-LTE effets due to high critical densities. 
CH$_3$OD transitions are also optically thick with opacities ranging from 0.3 to 1.3 towards the two sources.
The CHD$_2$OH population distribution shows much less scatter that CH$_3$OD.
The comparison of the column densities derived in this work with those found for CH$_3$OH in \citet{Taquet2015} allows us to derive the methanol deuterations in the two sources. We find a [CH$_2$DOH]/[CH$_3$OH] abundance ratio of $5.8 \pm 0.8$ and $3.2 \pm 0.7$ \% towards IRAS2A and IRAS4A, respectively. {Using the CH$_3$OD partition function estimated by \citet{Parise2004}, we find [CH$_2$DOH]/[CH$_3$OD] abundance ratios of $3.6 \pm 1.6$ and $5.3 \pm 3.4$ in IRAS2A and IRAS4A, respectively. However, as expected, we find lower [CH$_2$DOH]/[CH$_3$OD] abundance ratios of $0.8 \pm 0.3$ and $1.2 \pm 0.5$ in IRAS2A and IRAS4A, respectively, when the partition function from \citet{Jorgensen2018} is used. As discussed in Section \ref{comp_model}, a [CH$_2$DOH]/[CH$_3$OH] abundance ratio of $\sim 3$ is consistent with the statistical value whereas a ratio of $\sim 1$ remains puzzling. }
In addition to methanol, we also detect three transitions from the deuterated methyl cyanide isotopologue CH$_2$DCN at 225 GHz. We obtain the CH$_2$DCN column densities by assuming that the CH$_2$DCN rotational temperature is equal to 200 K, the rotational temperature of CH$_3$CN measured in \citet{Taquet2015}. Comparing the CH$_2$DCN column densities derived in this work with those of CH$_3$CN from \citet{Taquet2015} allows us to obtain [CH$_2$DCN]/[CH$_3$CN] ratios of $3.6 \pm 0.8$ and $2.7 \pm 0.9$ \% in IRAS2A and IRAS4A, respectively.

For HH212, we re-derive the column densities of $^{13}$CH$_3$OH and CH$_2$DOH already estimated by \citet{Bianchi2017}. Our methodology differs slightly with respect to \citet{Bianchi2017} since the PD diagram is built upon the flux extracted from an elliptical mask instead of the continuum peak position in order to use consistent methods for the different sources. Assuming a source size of 0$\farcs$19 following \citet{Bianchi2017}, we start with $^{13}$CH$_3$OH and we then use the derived rotational temperature to estimate the CH$_2$DOH and CH$_3$OD column densities.
We obtain a [CH$_2$DOH]/[CH$_3$OH] abundance ratio of $2.9 \pm 0.8$ \%. The [CH$_2$DOH]/[CH$_3$OD] abundance ratio is found to be $7.2 \pm 5.3$ and $1.5 \pm 0.6$ with the CH$_3$OD partition function estimated by \citet{Parise2004} and \citet{Jorgensen2018}, respectively.

\section{Discussion}

\subsection{Comparison with previous observations}

Deuterated methanol has been previously detected with the IRAM 30 meters single-dish telescope towards IRAS2A and IRAS4A. 
\citet{Parise2006} derived [CH$_2$DOH]/[CH$_3$OH] abundance ratios of 62 and 65 \% towards IRAS2A and IRAS4A, respectively, which are 10 and 17 times higher than the deuteration values found in this study.
The [CH$_3$OD]/[CH$_3$OH] and [CH$_2$DOH]/[CH$_3$OD] abundance ratios derived in this work highly depend on the used partition function (see Section \ref{spec_params}). Using the same partition function from \citet{Parise2004}, the [CH$_3$OD]/[CH$_3$OH] ratio of 4.7 \% found by \citet{Parise2006} towards IRAS4A is 6.7 times higher than our derived value 
{whilst the [CH$_2$DOH]/[CH$_3$OD] is found to decrease by a factor of 4 between \citet{Parise2006} and this work.}
For doubly-deuterated methanol, the [CHD$_2$OH]/[CH$_3$OH] ratios derived in this work are also much smaller than the abundances derived by \citet{Parise2006}, by a factor of 5.6 and 8.1 for IRAS2A and IRAS4A respectively. 

The large differences found between \citet{Parise2006} and the present work can be explained by several factors.
First, the new CH$_2$DOH spectroscopic data by \citet{Pearson2012} used in this work decreases by a factor of two the derived CH$_2$DOH column density following a re-analysis of the single dish data towards IRAS2A by \citet{Parise2006} using the most recent spectroscopy data entry. This also naturally explains the increase of the [CHD$_2$OH]/[CH$_2$DOH] ratio in this work by a factor of $\sim$ two, from 41 to 75 \% in IRAS2A and from 26 to 56 \% in IRAS4A. 
Second, the methanol deuteration derived from single-dish observations by \citet{Parise2006} has been estimated from the detection of various transitions from the main CH$_3$OH isotopologue. However, the targeted transitions have low upper level energies ($E_{\rm up} < 150$ K) and are optically thick in the inner protostellar regions surrounding these two low-mass protostars \citep[see][]{Taquet2015}. Therefore, the rotational diagram analysis, assuming optically thin emission, carried out in \citet{Parise2006} could have likely led to an underestimation of the CH$_3$OH column density. 
Finally, the IRAM 30 meters telescope has a large beam of 9-30$\arcsec$ depending on the frequency. As suggested by the low rotation temperatures of 55 and 27 K found by \citet{Parise2006} for CH$_2$DOH towards IRAS2A and IRAS4A, respectively, the methanol emission would mostly come from the large external envelope and/or the molecular outflows driven by the targeted sources where methanol deuteration is higher. 
Indeed, methanol is mostly produced through surface chemistry through CO hydrogenation as gas phase chemistry is known to be inefficient \citep{Watanabe2002, Geppert2006}. Non-thermal evaporation processes could release a fraction of solid methanol formed in ices into the gas phase. As shown by \citet{Taquet2014}, the deuteration of methanol and other molecules mostly produced in ices like water is higher in external protostellar envelopes than the inner hot cores since the cold gaseous deuteration reflects the deuteration at the highly deuterated ice surface. 

Deuterated methanol has not previously been targeted with single-dish facilities towards HH212. Our PD analysis gives a [CH$_2$DOH]/[CH$_3$OH] abundance ratio of $2.9 \pm 0.8$ \% in good agreement with the value of $2.4 \pm 0.4$ \% found by \citet{Bianchi2017} who used a slightly different method.

\subsection{Deuteration from low- to high-mass protostars}

\begin{table*}[htp]
\centering
\caption{Surrounding cloud dust temperatures and statistical deuterium fractionations observed towards
  protostellar hot-cores.}
\begin{footnotesize}
\begin{tabular}{l c c c c c c c}
\hline                                    
\hline                                                          
Source	&	$T_{\rm dust}$	&	D/H (CH$_2$DOH)	&	D/H (CH$_3$OD)	&	[CH$_2$DOH]	&	D/H (CH$_2$DCN)	&	D/H (HDO)	&	Ref.	\\
	&	(K)	&	(\%)	&	(\%)	&	/[CH$_3$OD]	&	(\%)	&	(\%)	&		\\
\hline															
IRAS 16293-B	&	16.6 $\pm$ 1.1	&	2.4 $\pm$ 0.5	&	1.8 $\pm$ 0.4	&	$3.9 \pm 1.1$	&	1.2 $\pm$ 0.3	&	0.046 $\pm$ 0.013	&	[1, 2, 3]	\\
IRAS 2A	&	13.1 $\pm$ 0.7	&	1.9 $\pm$ 0.5	&	$1.6 - 7.1$	&	$0.8-3.6$	&	1.2 $\pm$ 0.4	&	0.085 $\pm$ 0.040	&	[4, 5]	\\
IRAS 4A	&	13.1 $\pm$ 0.7	&	1.2 $\pm$ 0.4	&	$0.7 - 3.1$	&	$1.2-5.3$	&	0.90 $\pm$ 0.37	&	0.17 $\pm$ 0.08	&	[4, 6]	\\
HH212	&	15.3 $\pm$ 0.7	&	1.0 $\pm$ 0.3	&	$0.4 - 2.0$	&	$1.5-7.2$	&	-	&	-	&	[7, 4]	\\
NGC 7129 FIRS1	&	19.3 $\pm$ 0.2	&	0.21 $\pm$ 0.04	&	-	&	-	&	-	&	-	&	[8]	\\
Sgr B2	&	24 $\pm$ 1	&	0.04	&	0.07	&	1.7	&	0.13	&	-	&	[9]	\\
Orion KL	&	24.3 $\pm$ 0.4	&	0.037 $\pm$ 0.007	&	0.15 $\pm$ 0.05	&	$0.73 \pm 0.27$	&	-	&	0.13 $\pm$ 0.06	&	[10, 11]	\\
NGC 6334 IMM1I	&	23.9 $\pm$ 0.8	&	0.020 $\pm$ 0.007	&	0.48 $\pm$ 0.12	&	$0.13 \pm 0.03$	&	-	&	-	&	[12]	\\
\hline                                   
\end{tabular}
\label{table_Dmeth}
\tablebib{
{For the CH$_3$OD column densities in IRAS2A, IRAS4A, and HH212, two estimates are given using the partition functions estimated by \citet{Parise2004} and \citet{Jorgensen2018}. The highest CH$_3$OD column densities (and the lowest [CH$_2$DOH]/[CH$_3$OD] ratios) are those with the partition function estimated by \citet{Jorgensen2018}.}
[1]: \citet{Jorgensen2018};
[2]: \citet{Calcutt2018};
[3]: \citet{Persson2013};
[4]: This work;
[5]: \citet{Coutens2014};
[6]: Average between HDO/H$_2$O ratios of $5 \times 10^{-3}$ \citep{Taquet2013} and $5.4 \times 10^{-4}$ \citep[][M. Persson, priv. comm.]{Persson2014};
[7]: \citet{Bianchi2017};
[8]: \citet{Fuente2014};
[9]: \citet{Belloche2016};
[10]: \citet{Peng2012};
[11]: \citet{Neill2013};
[12]: \citet{Bogelund2018}
}
\end{footnotesize}
\end{table*}

\begin{figure*}[htp]
\centering
\includegraphics[width=\textwidth]{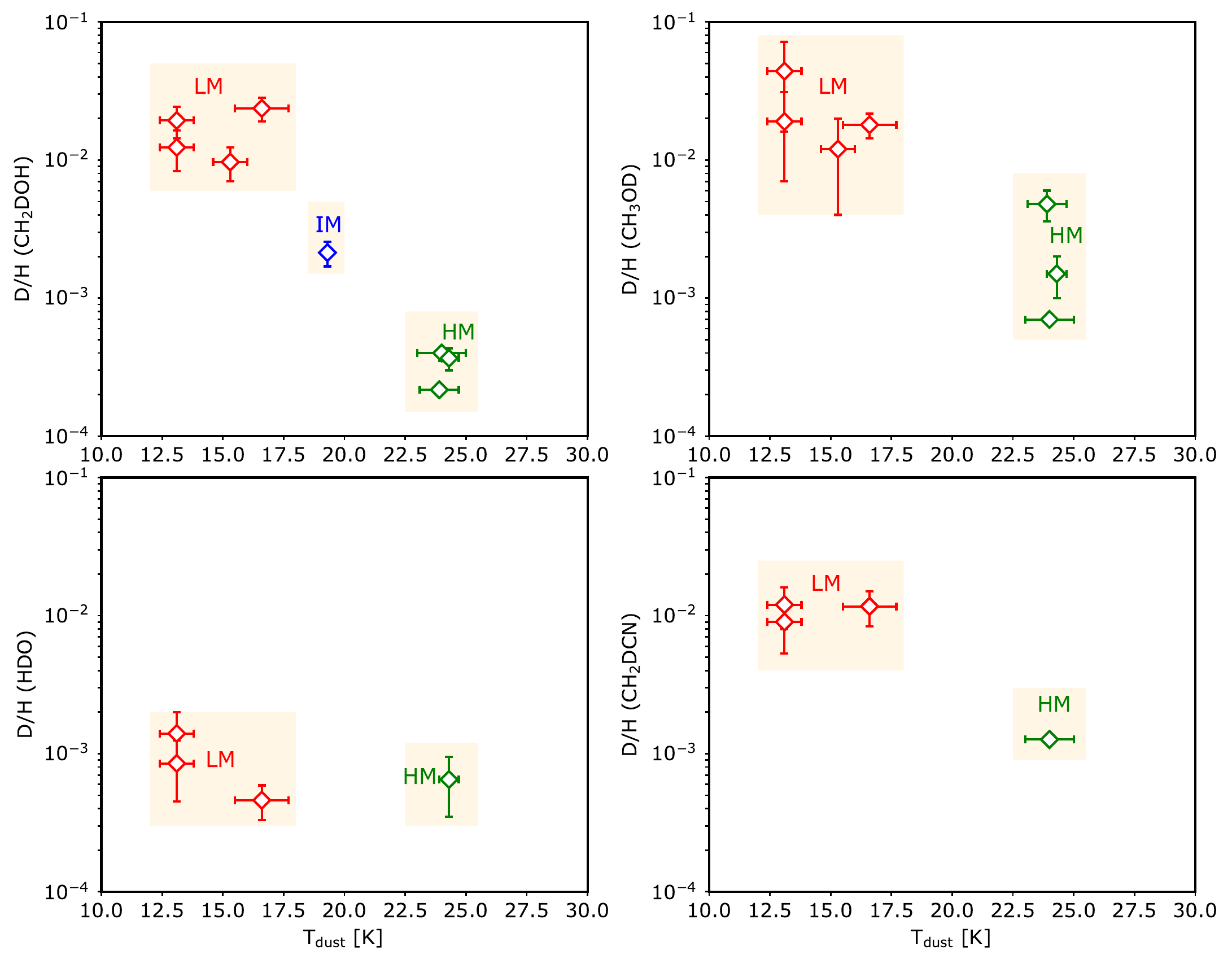} 
\caption{{Statistical {(i.e., taking into account the statistical ratios for the functional groups with multiple H-atoms)} CH$_2$DOH (top left), CH$_3$OD (top right), 
  HDO (bottom left), and CH$_2$DCN (bottom right) } deuterium fractionations observed towards
  the hot core of a sample of low-mass (red), intermediate-mass (blue), and  high-mass (green) protostars with sub-mm interferometers as function of the dust temperature of the surrounding cloud measured with the {\it Herschel} or {\it Planck} telescopes (see values in Table \ref{table_Dmeth}). 
  }
\label{ch2doh_obs_mod}
\end{figure*}

Figure \ref{ch2doh_obs_mod} and Table \ref{table_Dmeth} compare the deuteration of CH$_2$DOH, CH$_3$OD, and CH$_2$DCN, including their statistical {(i.e., taking into account the statistical ratios for the functional groups with multiple H-atoms)} correction, between four low-mass, one intermediate-mass, and three massive hot cores obtained only through interferometric observations of methanol isotopologues. We also show the water deuteration for comparison when available. 
Since the main CH$_3$OH isotopologue is known to be optically thick in hot cores, we only choose observations whose deuteration values were derived from $^{13}$CH$_3$OH or CH$_3^{18}$OH column densities {assuming $^{12}$C/$^{13}$C and $^{16}$O/$^{18}$O elemental ratios of 70 and 560, respectively \citep{Wilson1994}}.
IRAS 16293--2422, a low-mass protobinary system located in the Ophiuchus molecular cloud, has been observed with ALMA by \citet{Jorgensen2018} in the context of the PILS survey that covered the whole ALMA band 7 between between 329.15 and 362.90 GHz \citep{Jorgensen2016}.
NGC 7129 FIRS 2, an intermediate mass protostar, has been observed with the IRAM-PdBI at $220$ GHz by \citet{Fuente2014}.
The methanol deuterium fractionation towards the Orion KL massive star-forming region has been estimated by \citet{Peng2012} through interferometric observations using the IRAM-PdBI at $103$ and $225$ GHz.
The Sgr B2(N2) massive hot core has been observed by \citet{Belloche2016} in the context of the EMoCA ALMA survey of the whole 3 mm band.
The NGC6334 massive protocluster has recently been observed by \citet{Bogelund2018} through ALMA observations at $300$ GHz and we selected the deuteration derived using the CH$_3^{18}$OH column density.
Note that the error bars of the CH$_3$OD deuterations derived in this work include the different column density values derived with the two partition functions used for CH$_3$OD. 

The dust temperature in the dark clouds surrounding protostars is mostly governed by the external interstellar radiation field with local variations due to bright stars and/or dense cores. All the hot cores considered in Figure \ref{ch2doh_obs_mod} and Table \ref{table_Dmeth} are likely relatively young, i.e. with a lifetime of $10^5$ yr at most, the dust temperatures of their precursor dark clouds are therefore likely similar to the current observed values.
The deuterium fractionation is plotted against the current dust temperature of their surrounding cloud as measured by the {\it Planck} observatory for NGC 7129 FIRS2 and by the PACS and SPIRE instruments onboard the {\it Herschel Space Observatory} for other sources. The Ophiuchus, Perseus, and Orion molecular clouds have been observed with the Gould Belt Survey key program \citep{Andre2010}, the Sgr B2 region has been observed by \citet{Etxaluze2013} as part of the Hi-GAL key program \citep{Molinari2010}, whilst the NGC6334 massive complex has been observed with the HOBYS key program \citep{Motte2010, Russeil2013, Tige2017}. 
To estimate the dust temperature, we extract 10 arcmin maps surrounding the selected sources. We then generate temperature histograms of all pixels showing visual extinctions $A_{\rm V}$ higher than 3 mag when the H$_2$ column density maps are available (for the Ophiuchus, Perseus, and NGC6334 clouds) or showing 353 GHz opacities higher than 10 times the rms noise (for the Orion and NGC 7129 clouds). To derive a temperature representative of the entire region, we filter out the "hot" regions locally heated by stars or "cold" regions of dense cores by fitting the temperature histogram arround the main peak with a gaussian function. Figure \ref{temp_cloud} of the Appendix gives an example for the NGC1333 region containing IRAS2A and IRAS4A.

It is found that the statistical deuterium fractionations of CH$_2$DOH and CH$_3$OD tend to decrease from a few percents to less than $\sim$ 0.1 \% with the dust temperature between $\sim 13$ and $\sim 24$ K. 
{Using the same partition function estimated by \citet{Belloche2016} and \citet{Jorgensen2018}, the [CH$_2$DOH]/[CH$_3$OD] abundance ratio seems to show a large dispersion since it typically varies from $1-4$ in low-mass protostars to $0.1-2$ in high-mass hot cores ($0.8 \pm 0.3$ in IRAS2A, $1.2 \pm 0.5$ in IRAS4A, $3.9 \pm 1.1$ in IRAS16293, $1.5 \pm 0.6$ in HH212, 1.7 in Sgr B2, $0.73 \pm 0.27$ in Orion KL, and $0.13 \pm 0.03$ in NGC6334-IMM1). }

It has been shown that warm "hot-core" gas phase chemistry is likely too slow to significantly alter the deuteration after the evaporation of ices in hot cores \citep{Charnley1997, Osamura2004}. 
Thus, the methanol deuterium fractionation observed in the warm gas phase of low-mass and massive hot cores would rather reflect the deuteration of methanol formed in interstellar ices within their progenitor dense clouds.

\subsection{Comparison with theoretical predictions} \label{comp_model}

We compare the observed methanol deuterations with the predictions of the GRAINOBLE astrochemical code by \citet{Taquet2012, Taquet2013, Taquet2014} which follows the formation and the deuteration of ices in cold dense cores. The code has been extensively described in previous studies and has recently been used to interpret the deuteration towards NGC6334 measured by \citet{Bogelund2018}. 
We investigate the effect of the temperature, total density, and time on the methanol deuteration using the same code and same chemical network than in \citet{Taquet2014}. However, unlike in \citet{Taquet2014}, we run here a series of pseudo-time dependent simulations in which the chemistry evolves over time for constant physical properties such as temperature or density. 
Figure \ref{deut_obs_mod} {of the Appendix} shows the deuteration of methanol and water in ices as a function of temperature between 10 and 30 K for three different dense cloud densities, $n_{\rm H} = 10^4$, $10^5$ and $10^6$ cm$^{-3}$ at three different times, $0.1 \times t_{\rm FF}$, $t_{\rm FF}$, and $10 \times t_{\rm FF}$, where $t_{\rm FF}$ is the free-fall time. $t_{\rm FF}$ is equal to $4.4 \times 10^5$, $1.4 \times 10^5$ and $4.4 \times 10^4$ yr at densities of $10^4$, $10^5$ and $10^6$ cm$^{-3}$, respectively. 

The methanol deuteration strongly depends on the density, the considered temperature, and time. Methanol and its deuterated isotopologues are formed through addition reactions of atomic H and D on CO and H$_2$CO on ices possibly supplemented by abstraction and {substitution} reactions \citep{Nagaoka2005, Hidaka2007, Hidaka2009}. 
The methanol deuteration is therefore governed by the atomic [D]/[H] abundance ratio in the gas phase during ice formation. Atomic D is mostly formed via electronic recombination of H$_3^{+}$ isotopologues, formed through exothermic reactions between H$_3^{+}$ and HD. The efficiency of the backward reaction increases with the increasing temperature and with the ortho/para ratio of H$_2$. In addition, reactions between H$_3^{+}$ or with its isotopologues and HD are in competition with reactions involving CO. 
The production of atomic deuterium is therefore enhanced at low temperatures, close to 10 K, when the abundances of CO and ortho state of H$_2$ are low \citep[see Fig. 1 of][for a scheme detailing the gas phase deuteration network]{Taquet2012b}. 
As a consequence, the strong decrease, up to two orders of magnitudes, of the methanol deuteration predicted at $t \geq t_{\rm FF}$ from 10 to 30 K is essentially due to the decreased efficiency of deuterium chemistry in the gas phase of cold dense cores. 

The observed [CH$_3$OD]/[CH$_3$OH] ratios observed both towards low-mass and massive protostars are relatively well reproduced by the model predictions at $t \geq t_{\rm FF}$, in spite of the large dispersion of values among massive protostars. The [CH$_2$DOH]/[CH$_3$OH] abundance ratios of low-mass protostars can also be explained by the models at $t \geq t_{\rm FF}$ but the massive protostar values can only be explained by a shorter time of $0.1 \times t_{\rm FF}$ or a low density of $10^4$ cm$^{-3}$.
{The predicted [CH$_2$DOH]/[CH$_3$OD] ratio remains close the statistical value of 3 in most cases and can increase up to $\sim 6$ at $10^6$ cm$^{-3}$, $T \sim 10-15$ K, and $t = 10 \times t_{\rm FF}$. 
Our model only focusing on cold surface chemistry is not able to explain [CH$_2$DOH]/[CH$_3$OD] ratios lower than 1 observed in massive hot cores. CH$_3$OD would then need to be produced through other processes such as warm gas phase chemistry following methanol evaporation within massive hot cores. \citet{Osamura2004} showed that the reaction H$_2$DO$^+$ + CH$_3$OH $\rightarrow$ CH$_3$OHD$^+$ + H$_2$O, followed by dissociative recombination, could increase the [CH$_3$OD]/[CH$_3$OH] ratio by a factor of $\sim 5$ at 100 K. However, such an increase is obtained at times longer than $10^5$ yr and assuming [HDO]/[H$_2$O] abundance ratios of $\sim 0.1$ that are about 100 times larger than the observed values. It remains to be tested whether this process could play a major role at higher temperatures and for realistic water deuteration values. 
On the other hand, hydrogen-deuterium exchanges in warm ices have been proven to be efficient but only on the hydroxyl functional group of methanol. CH$_3$OD would thus give its deuterium to water, inducing a decrease of the CH$_3$OD abundance in ices before its evaporation \citep{Ratajczak2009, Faure2015}.
}

Unlike methanol, water does not show any significant decrease of its observed deuteration with the dust temperature since it remains around 0.1 \% between 13 and 24 K. The low water deuteration has been interpreted as due to an early formation of solid water in the translucent phase at low visual extinctions, when the dark cloud is still lukewarm, with temperatures of $15-20$ K, and when the abundances of CO and ortho H$_2$ are still high \citep{Taquet2013, Taquet2014}. 
Methyl cyanide is thought to be mostly produced either on warm ices from recombination between the CN and CH$_3$ radicals or in the gas phase through the radiative association between HCN and CH$_3^+$, after the prestellar stage. Comparing the methyl cyanide deuteration with the model predictions with constant physical conditions of cold cores is therefore not relevant.

\subsection{Deuteration from low-mass protostars to comets}

\begin{figure*}[htp]
\centering
\includegraphics[width=\textwidth]{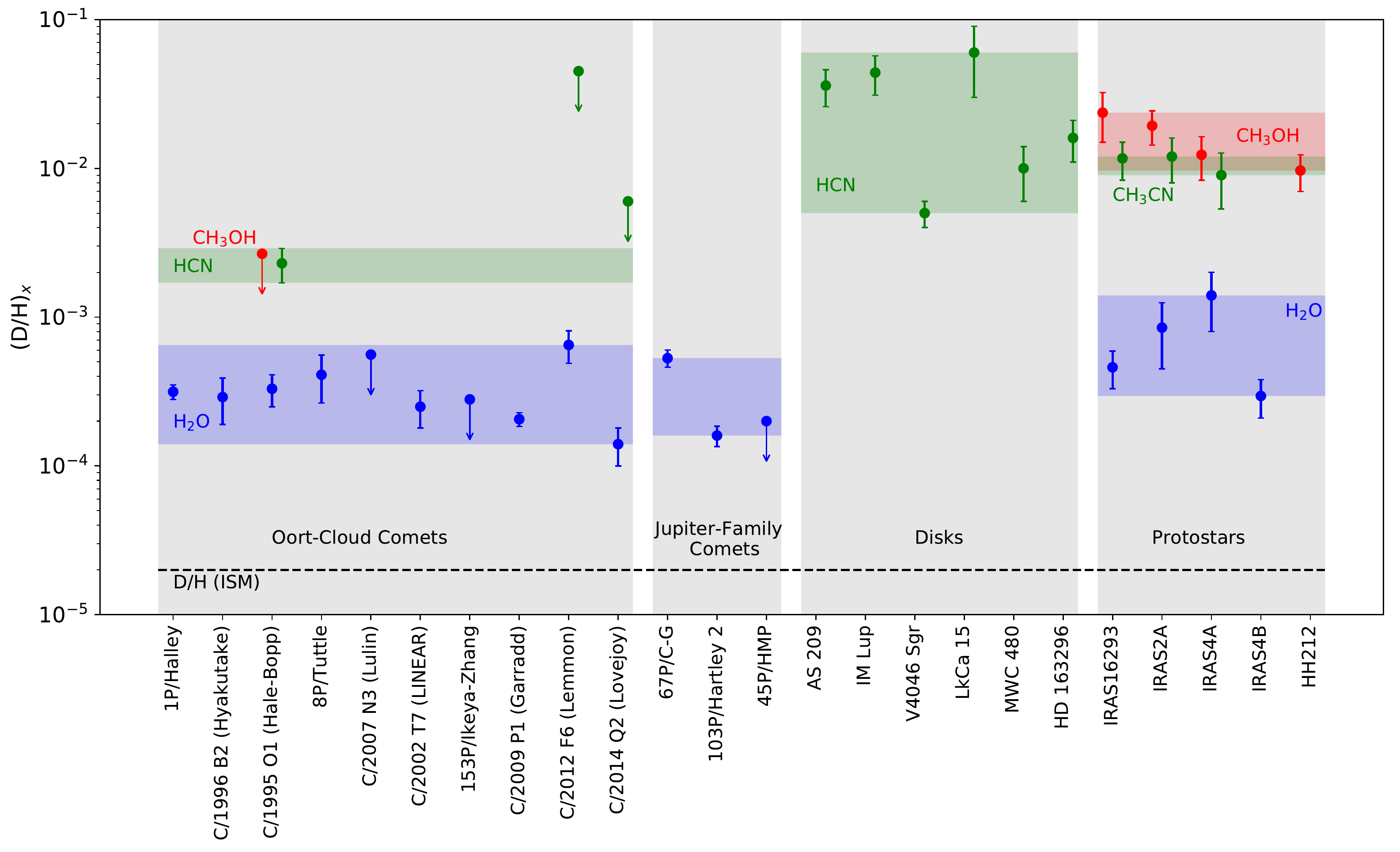} 
\caption{Statistical D/H ratios in Solar System comets, protoplanetary disks, and low-mass Class 0 protostars for methanol (red), water (blue), and HCN or CH$_3$CN (green). 
{\bf References.}
1P/Halley: \citet{Eberhardt1995};
C/1996 B2 (Hyakutake): \citet{BockeleeMorvan1998};
C/1995 O1 (Hale-Bopp): \citet{Meier1998, Crovisier2004};
8P/Tuttle: \citet{Villanueva2009};
C/2007 N3 (Lulin): \citet{Gibb2012};
C/2002 T7 (LINEAR): \citet{Hutsemekers2008};
153P/Ikeya-Zhang: \citet{Biver2006};
C/2009 P1 (Garradd): \citet{BockeleeMorvan2012};
C/2012 F6 (Lemmon): \citet{Biver2016};
C/2014 Q2 (Lovejoy): \citet{Biver2016};
67P/C-G: \citet{Altwegg2015};
103P/Hartley 2: \citet{Hartogh2011};
45P/HMP: \citet{Lis2013};
All disks: \citet{Huang2017};
IRAS 16293: \citet{Persson2014, Jorgensen2018, Calcutt2018};
IRAS2A: \citet{Coutens2014}, this work;
IRAS4A: Average of values derived by \citet{Taquet2013b} and \citet[][M. Persson priv. comm.]{Persson2014}, this work;
IRAS4B: \citet{Persson2014}.
}
\label{deut_proto_comets}
\end{figure*}

Solar System comets mostly contain pristine material that has poorly evolved since the end of the Solar System formation 4.5 billions years ago. Their chemical composition should reflect the chemical composition of the solar nebula \citep[see][for instance]{Mumma2011}. 
{Comparing the deuteration of water and organics measured around young solar-type protostars on Solar System scales and in Solar System comets should allow us to follow the chemical evolution throughout the star formation process. }

Figure \ref{deut_proto_comets} compares the (statistical) deuterium fractionation of methanol, methyl cyanide, and water observed in low-mass protostars with the deuteration of HCN in protoplanetary disks and of CH$_3$OH, HCN, and water in comets. Deuteration of water has been estimated in four nearby protostars on Solar System scales by \citet{Taquet2013, Persson2014} and \citet{Coutens2014} through interferometric observations of HDO and H$_2^{18}$O transitions and about a dozen of comets (see the list of references in caption of Fig. \ref{ch2doh_obs_mod}). 
Water deuteration in protostars is only higher roughly by a factor of two than the deuteration measured in comets. 
Unlike water, cometary deuterated methanol has not been detected so far. The most stringent upper limit in the methanol deuteration has been given by \citet{Crovisier2004} towards comet Hale-Bopp, with a [CH$_2$DOH]/[CH$_3$OH] abundance ratio lower than 0.8\%. The [CH$_2$DOH]/[CH$_3$OH] ratio of $3-6$ \% estimated with interferometers towards young protostars is higher by a factor of $4-8$ than the estimate made in comet Hale-Bopp. 
A low methanol deuteration in comets remains to be confirmed with a clear detection of deuterated methanol, either through the analysis of the high resolution ROSINA mass spectrometer data at mass 33 onboard the {\it Rosetta} space probe \citep[see for instance][]{LeRoy2015}, or through future sensitive sub-mm/mm ground-based surveys of nearby and bright comets using interferometers like ALMA or NOEMA.


Assuming that the observed deuteration towards nearby low-mass protostars is representative to the deuteration of the young Sun, this comparison suggests that an important reprocessing of the organic material occured in the solar nebula whilst little water reprocessing in the solar nebula is required.
One can suggest three reasons.  
First, methanol observed in the hot core of protostars on $\sim 100$ AU scales is reprocessed before entering in the disk midplane, through high-temperature chemistry involving reactions with H or OH {since they have moderate energy barriers of 2200-3000 K \citep{Li1996}}. However, the temperature of the hot corino is likely too low to trigger such reactions and other processes that increase the gas temperature to $\sim 1000$ K, such as accretion shocks that form at the interface of the envelope and the disk in formation \citep{Aota2015}, would be needed.
Second, physical processes, such as turbulence, in the disk could transport the material from the disk midplane to the atmosphere where efficient photolytic processes can photodissociate organic molecules. \citet{Albertsson2013} and \citet{Furuya2013} demonstrated that radial and vertical turbulence can decrease the water deuteration by typically one order of magnitude in the cometary zone. However, it is yet to be demonstrated that this process could also impact the deuteration of more complex organic species with slightly different chemistries.
A third possibility is that the cloud at the origin of the Solar System was initially slightly warmer than the temperatures of nearby dark clouds in which the observed protostars are located. This has been recently suggested by \citet{Taquet2016} who concluded that a dark cloud temperature of $\sim 20$ K, together with a dark cloud density higher than $10^5$ cm$^{-3}$, was needed to explain the high abundance of O$_2$ and its strong correlation with water in comet 67P/C-G. {This modelling work confirmed the studies of meteoretic data which suggest that the Solar System was born in a dense cluster of stars \citep[see][]{Adams2010}}. According to Fig. \ref{ch2doh_obs_mod}, an increase of the dark cloud temperature from 10 to $20$ K would decrease the methanol deuteration from $2-6$ to $0.2 - 0.6$ \%, depending on the density, which seems to be in good agreement with the value found in comet 67P/C-G.

\section{Conclusions}

We analysed several existing observational datasets obtained with the PdBI and ALMA sub-mm interferometers to estimate the methanol deuterations in the hot cores surrounding three protostars on Solar System scales. For this purpose, we analysed several dozens of deuterated methanol transitions with a Population Diagram analysis in order to measure the [CH$_2$DOH]/[CH$_3$OH] and [CH$_3$OD]/[CH$_3$OH] abundance ratios. 
The obtained [CH$_2$DOH]/[CH$_3$OH] ratios of $3-6$ \% and [CH$_3$OD]/[CH$_3$OH] ratios of $0.4-1.6$ \% are typically one order of magnitude lower than previous estimates derived from single-dish observations towards the same sources, and are in good agreement with the recent ALMA measurements by \citet{Jorgensen2018} towards the low-mass protostar IRAS16293-B. 
We then compared our methanol deuteration estimates with previous measurements of intermediate- and high-mass hot cores with similar observational properties and analysis methods. Unlike water which does not show strong variation of its deuteration between low-mass and high-mass protostars, we find that the methanol deuteration around massive hot cores is much lower by one to two orders of magnitude. This strong difference could be attributed to a different physical/chemical history of the sources. Methanol observed around protostars is mostly formed in interstellar ices during the previous molecular cloud phase. Dust temperature maps derived with the {\it Herschel} or Planck space observatories suggest that the observed massive protostars were born in warm molecular clouds of $T \sim 24$ K whilst low-mass protostars are located in more quiescent and cold regions with temperatures lower than 15 K. Comparing the observed deuteration values with the predictions of the \texttt{GRAINOBLE} astrochemical model on ice formation and deuteration, an increase of 10 K in the dust temperature is enough to explain a decrease of the observed methanol deuteration, depending on the density and the chemical time. 
Finally, the methanol deuterations measured towards young solar-type protostars at high angular resolution on Solar System scales seem to be higher by a factor of $\sim 5$ than the upper limit in methanol deuteration estimated in comet Hale-Bopp by \citet{Crovisier2004}. If this result is confirmed by subsequent observations of other comets, this would imply that an important reprocessing of the organic material likely occurred in the solar nebula during the formation of the Solar System.

\begin{acknowledgements}

This work is based on observations carried out with the ALMA Interferometer under project numbers ADS/JAO.ALMA\#2012.1.00997.S and ADS/JAO.ALMA\#2016.1.01475.S data (PI: C. Codella) and with the IRAM PdBI/NOEMA Interferometer under project numbers V05B and V010 (PI: M.V. Persson) and U003 (PI: V. Taquet). 
ALMA is a partnership of ESO (representing its member states), NSF (USA) and NINS (Japan), together with NRC (Canada) and NSC and ASIAA (Taiwan), in cooperation with the Republic of Chile. The Joint ALMA Observatory is operated by ESO, AUI/NRAO and NAOJ. 
IRAM is supported by INSU/CNRS (France), MPG (Germany) and IGN (Spain).
V.T. acknowledges the financial support from the European Union's Horizon 2020 research and innovation programme under the Marie Sklodowska-Curie grant agreement n. 664931.
E.B., Ce.Ce., C.K., A.L. and Cl.Co acknowledge the funding from the European Research Council (ERC) under the European Union's Horizon 2020 research and innovation programme, for the Project “The Dawn of Organic Chemistry” (DOC), grant agreement No 741002.
Cl.Co acknowledges the funding from PRIN-INAF 2016 "The Cradle of Life - GENESIS-SKA (General Conditions in Early Planetary Systems for the rise of life with SKA)”.
Ce.Ce. and Cl.Co acknowledge the financial support from the European MARIE SKŁODOWSKA-CURIE ACTIONS under the European Union's Horizon 2020 research and innovation programme, for the Project “Astro-Chemistry Origins” (ACO), Grant No 811312. 

\end{acknowledgements}

\bibliographystyle{aa}

\appendix

\section{Line parameters of the targeted transitions}

\begin{table*}[htp]
\centering
\caption{Line parameters of CH$_2$DOH lines observed towards IRAS2A and IRAS4A}
\begin{tabular}{l c c c c c c}
\hline                                    
\hline   
Number  & Frequency & Transition  & E$_{up}$  & A$_{ul}$  & Flux (IRAS2A) & Flux (IRAS4A) \\
  & (GHz) &   & (K) & (s$^{-1}$)  & (Jy km/s) & (Jy km/s) \\
\hline                                  
1 & 166.063164  & $2_{2,0} e_1 - 1_{1,0} o_1$ & 33.0  & 2.40(-5)  & 0.285 $\pm$ 0.081 & 0.199 $\pm$ 0.076 \\
2 & 166.787448  & $2_{2,1} e_1 - 1_{1,1} o_1$ & 33.0  & 2.43(-5)  & 0.292 $\pm$ 0.071 & 0.184 $\pm$ 0.051 \\
3 & 225.878232  & $3_{1,3} o_1 - 2_{0,2} o_1$ & 35.6  & 3.23(-5)  & 0.607 $\pm$ 0.123 & 0.204 $\pm$ 0.056 \\
4 & 226.818248  & $5_{1,4} e_0 - 4_{1,3} e_0$ & 36.7  & 3.58(-5)  & 0.931 $\pm$ 0.187 & 0.349 $\pm$ 0.074 \\
5 & 144.762399  & $4_{1,3} e_1 - 3_{0,3} o_1$ & 38.2  & 8.80(-6)  & 0.132 $\pm$ 0.028 & 0.153 $\pm$ 0.043 \\
6 & 164.108467  & $5_{0,5} e_1 - 4_{1,4} e_1$ & 45.6  & 2.67(-6)  & 0.086 $\pm$ 0.031 & 0.090 $\pm$ 0.031 \\
7 & 223.898819  & $3_{2,2} o_1 - 4_{2,2} e_1$ & 48.3  & 1.56(-6)  & 0.450 $\pm$ 0.095 & 0.219 $\pm$ 0.068 \\
8 & 225.667709  & $5_{1,4} e_1 - 4_{1,3} e_1$ & 49.0  & 4.44(-5)  & 0.272 $\pm$ 0.062 & 0.351 $\pm$ 0.075 \\
9 & 224.928016  & $5_{1,4} e_2 - 4_{1,3} e_2$ & 55.3  & 4.42(-5)  & 1.390 $\pm$ 0.280 & 0.428 $\pm$ 0.092 \\
10  & 144.134719  & $5_{3,2} e_0 - 6_{2,5} e_0$ & 68.1  & 2.64(-6)  & 0.069 $\pm$ 0.026 & 0.098 $\pm$ 0.035 \\
11  & 223.691457  & $5_{3,3} e_0 - 4_{3,2} e_0$ & 68.1  & 2.17(-5)  & 0.601 $\pm$ 0.122 & 0.203 $\pm$ 0.043 \\
12  & 223.697110  & $5_{3,2} e_0 - 4_{3,1} e_0$ & 68.1  & 2.17(-5)  & 0.608 $\pm$ 0.124 & 0.217 $\pm$ 0.051 \\
13  & 225.848108  & $7_{0,7} o_1 - 6_{1,6} o_1$ & 78.3  & 2.17(-5)  & 0.604 $\pm$ 0.123 & 0.540 $\pm$ 0.114 \\
14  & 165.861895  & $8_{0,8} e_1 - 8{1,7} e_0$  & 90.4  & 8.99(-6)  & 0.217 $\pm$ 0.057 & 0.056 $\pm$ 0.020 \\
15  & 164.577869  & $8_{1,7} e_1 - 7_{2,6} e_1$ & 94.4  & 1.52(-6)  & 0.058 $\pm$ 0.023 & 0.071 $\pm$ 0.029 \\
16  & 223.616196  & $5_{4,2} e_0 - 4_{4,1} e_0$ & 95.1  & 1.26(-5)  & 0.522 $\pm$ 0.107 & 0.264 $\pm$ 0.065 \\
  & 223.616210  & $5_{4,1} e_0 - 4_{4,0} e_0$ & 95.1  & 1.26(-5)  &       &       \\
17  & 164.374775  & $10_{4,7} e_0 - 10_{3,7} o_1$ & 181.0 & 5.91(-6)  & 0.073 $\pm$ 0.029 & 0.062 $\pm$ 0.031 \\
18  & 164.388327  & $10_{4,6} e_0 - 10_{3,8} o_1$ & 181.0 & 5.91(-6)  & 0.087 $\pm$ 0.032 & 0.106 $\pm$ 0.043 \\
19  & 143.566661  & $12_{2,10} e_0 - 11_{3,9} e_0$  & 184.5 & 3.76(-6)  & 0.082 $\pm$ 0.027 & 0.041 $\pm$ 0.016 \\
20  & 224.273919  & $10_{5,6} e_0 - 11_{4,7} e_0$ & 215.4 & 1.16(-5)  & 0.126 $\pm$ 0.035 & 0.082 $\pm$ 0.025 \\
21  & 224.285457  & $10_{5,5} e_0 - 11_{4,8} e_0$ & 215.4 & 1.16(-5)  & 0.178 $\pm$ 0.041 & 0.152 $\pm$ 0.038 \\
22  & 166.950015  & $12_{4,9} e_0 - 12_{3,9} o_1$ & 230.4 & 6.80(-6)  & 0.079 $\pm$ 0.031 & 0.065 $\pm$ 0.018 \\
23  & 166.995876  & $12_{4,8} e_0 - 12_{3,10} o_1$  & 230.4 & 6.80(-6)  & 0.072 $\pm$ 0.027 & 0.064 $\pm$ 0.021 \\
24  & 142.838827  & $16_{1,15} o1 - 16_{0,16} o1$ & 316.2 & 9.85(-6)  & 0.138 $\pm$ 0.031 & 0.099 $\pm$ 0.026 \\
25  & 225.551640  & $17_{2,15} o1 - 17_{1,16} o1$ & 363.6 & 3.28(-5)  & 0.273 $\pm$ 0.059 & 0.149 $\pm$ 0.042 \\
\hline                                  
\end{tabular}
\label{lines_ch2doh_2a4a}
\end{table*}

\begin{table*}[htp]
\centering
\caption{Line parameters of CH$_3$OD lines observed towards IRAS2A and IRAS4A}
\begin{tabular}{l c c c c c c}
\hline                                    
\hline   
Number  & Frequency & Transition  & E$_{up}$  & A$_{ul}$  & Flux (IRAS2A) & Flux (IRAS4A) \\
  & (GHz) &   & (K) & (s$^{-1}$)  & (Jy km/s) & (Jy km/s) \\
\hline                                  
1 & 226.53867 & $5_{0+} - 4_{0 +}$ A  & 32.7  & 4.31(-5)  & 0.869 $\pm$ 0.175 & 0.290 $\pm$ 0.064 \\
2 & 226.350191  & $5_0 - 4_0$ E & 36.4  & 4.66(-5)  & 0.744 $\pm$ 0.151 & 0.380 $\pm$ 0.084 \\
3 & 226.185930  & $5_{-1} - 4_{-1}$ E & 37.3  & 4.16(-5)  & 0.433 $\pm$ 0.088 & 0.206 $\pm$ 0.051 \\
4 & 143.741650  & $5_{1-} - 5_{0+}$ A & 39.6  & 3.52(-5)  & 0.237 $\pm$ 0.054 & 0.121 $\pm$ 0.029 \\
5 & 226.92258 & $5_{-2} - 4_{-2}$ E & 50.2  & 3.71(-5)  & 0.355 $\pm$ 0.074 & 0.143 $\pm$ 0.035 \\
6 & 226.892864  & $5_2 - 4_2$ E & 54.4  & 3.71(-5)  & 0.267 $\pm$ 0.055 & 0.113 $\pm$ 0.025 \\
7 & 226.94283 & $5_{2+} - 4_{2+}$ A & 54.4  & 3.70(-5)  & 0.436 $\pm$ 0.091 & 0.253 $\pm$ 0.067 \\
8 & 226.825536  & $5_3 - 4_3$ E & 70.5  & 3.71(-5)  & 0.254 $\pm$ 0.053 & 0.119 $\pm$ 0.026 \\
9 & 226.70660 & $5_{2-} - 4_{2-}$ A & 100 & 2.84(-5)  & 0.300 $\pm$ 0.062 & 0.149 $\pm$ 0.040 \\
10  & 226.73886 & $5_{-4} - 4_{-4}$ E & 104.3 & 1.60(-5)  & 0.073 $\pm$ 0.022 & 0.062 $\pm$ 0.031 \\
\hline                                  
\end{tabular}
\label{lines_ch3od_2a4a}
\end{table*}

\begin{table*}[htp]
\centering
\caption{Line parameters of CHD$_2$OH lines observed towards IRAS2A and IRAS4A}
\begin{tabular}{l c c c c c c}
\hline                                    
\hline   
Number  & Frequency & Transition  & E$_{up}$  & A$_{ul}$  & Flux (IRAS2A) & Flux (IRAS4A) \\
  & (GHz) &   & (K) & (s$^{-1}$)  & (Jy km/s) & (Jy km/s) \\
\hline                                  
1 & 166.435 & $4_0 - 3_0 e_0$ & 20.0  & 1.67(-5)  & 0.245 $\pm$ 0.066 & 0.151 $\pm$ 0.028 \\
2 & 166.327 & $4_0 - 3_0 o_1$ & 28.8  & 1.67(-5)  & 0.248 $\pm$ 0.107 & 0.129 $\pm$ 0.030 \\
3 & 166.234 & $4_0 - 3_0 e_1$ & 38.4  & 1.66(-5)  & 0.272 $\pm$ 0.064 & 0.139 $\pm$ 0.033 \\
4 & 166.271 & $4_{2-} - 3_{2-} e_1$ & 51.3  & 1.25(-5)  & 0.184 $\pm$ 0.066 & 0.106 $\pm$ 0.026 \\
5 & 166.304 & $4_{2+} - 3_{2+} e_1$ & 51.3  & 1.25(-5)  & 0.147 $\pm$ 0.045 & 0.074 $\pm$ 0.021 \\
6 & 166.297 & $4_{3+} - 3_{3-} e_1$ & 67.0  & 7.14(-6)  & 0.176 $\pm$ 0.055 & 0.048 $\pm$ 0.020 \\
  & 166.298 & $4_{3-} - 3_{3-} e_1$ & 67.0  & 7.14(-6)  &       &       \\
\hline                                  
\end{tabular}
\label{lines_chd2oh_2a4a}
\end{table*}

\begin{table*}[htp]
\centering
\caption{Line parameters of CH$_2$DCN lines observed towards IRAS2A and IRAS4A}
\begin{tabular}{l c c c c c c}
\hline                                    
\hline   
Number	&	Frequency	&	Transition	&	E$_{up}$	&	A$_{ul}$	&	Flux (IRAS2A)	&	Flux (IRAS4A)	\\
	&	(GHz)	&		&	(K)	&	(s$^{-1}$)	&	(Jy km/s)	&	(Jy km/s)	\\
\hline																	
1	&	224.754530	&	$13_{1,13}-12_{1,12}$ 	&	80.9	&	9.72(-4)	&	0.159	$\pm$	0.032	&	0.081	$\pm$	0.030	\\
2	&	225.723769	&	$13_{3,11}-13_{3,10}$ 	&	124.4	&	9.38(-4)	&	0.130	$\pm$	0.030	&	0.089	$\pm$	0.018	\\
3	&	225.724053	&	$13_{3,10}-13_{3,9}$ 	&	124.4	&	9.38(-4)	&	0.130	$\pm$	0.030	&	0.089	$\pm$	0.018	\\
4	&	225.726540	&	$13_{3,12}-13_{3,11}$ 	&	97.4	&	9.68(-4)	&	0.162	$\pm$	0.037	&	0.069	$\pm$	0.018	\\
5	&	225.781540	&	$13_{2,11}-13_{2,10}$ 	&	97.4	&	9.98(-4)	&	0.162	$\pm$	0.037	&	0.069	$\pm$	0.020	\\
\hline                                  
\end{tabular}
\label{lines_ch2dcn_2a4a}
\end{table*}

\begin{table*}[htp]
\centering
\caption{Line parameters of CH$_2$DOH lines observed towards HH212}
\begin{tabular}{l c c c c c c}
\hline                                    
\hline   
Number  & Frequency & Transition  & E$_{up}$  & A$_{ul}$  & Flux (HH212)  \\
  & (GHz) &   & (K) & (s$^{-1}$)  & (Jy km/s) \\
\hline                          
1 & 348.16076 & $4_{1,3} e_1 - 4_{0,4} e_0$ & 38.1  & 2.03(-4)  & 0.152 $\pm$ 0.031 \\
2 & 338.95711 & $6_{1,6} e_0 - 5_{0,5} e_0$ & 48.4  & 1.59(-4)  & 0.157 $\pm$ 0.031 \\
3 & 337.34866 & $9_{0,9} e_0 - 8_{1,8} o_0$ & 96.3  & 1.46(-4)  & 0.159 $\pm$ 0.032 \\
4 & 350.09024 & $5_{4,2} e_1 - 5_{3,2} o_1$ & 104.2 & 7.58(-5)  & 0.065 $\pm$ 0.065 \\
5 & 350.09038 & $5_{4,1} e_1 - 5_{3,3} o_1$ & 104.2 & 7.58(-5)  & 0.065 $\pm$ 0.065 \\
6 & 350.02735 & $6_{4,3} e_1 - 6_{3,3} o_1$ & 117.1 & 9.06(-5)  & 0.078 $\pm$ 0.016 \\
7 & 350.02777 & $6_{4,2} e_1 - 6_{3,4} o_1$ & 117.1 & 9.06(-5)  & 0.078 $\pm$ 0.016 \\
8 & 349.95168 & $7_{4,4} e_1 - 7_{3,4} o_1$ & 132.1 & 1.00(-4)  & 0.201 $\pm$ 0.040 \\
9 & 338.86898 & $13_{1,12} e_0 - 12_{0,12} o_1$ & 201.8 & 2.81(-5)  & 0.080 $\pm$ 0.016 \\
10  & 347.76728 & $18_{4,15} e_1 - 18_{3,15} o_1$ & 438.2 & 1.31(-4)  & 0.046 $\pm$ 0.009 \\
11  & 347.95281 & $18_{4,14} e_1 - 18_{3,16} o_1$ & 438.2 & 1.31(-4)  & 0.057 $\pm$ 0.011 \\
\hline                                  
\end{tabular}
\label{lines_ch2doh_hh212}
\end{table*}

\begin{table*}[htp]
\centering
\caption{Line parameters of the CH$_3$OD line observed towards HH212}
\begin{tabular}{l c c c c c c}
\hline                                    
\hline   
Number  & Frequency & Transition  & E$_{up}$  & A$_{ul}$  & Flux (HH212)  \\
  & (GHz) &   & (K) & (s$^{-1}$)  & (Jy km/s) \\
\hline                          
1 & 335.089661 & $6_{2-} - 6_{1+}$ & 67.4  & 1.65(-4)  & 0.025 $\pm$ 0.003 \\
\hline                                  
\end{tabular}
\label{lines_ch3od_hh212}
\end{table*}

\section{Column density and dust temperature maps}

\begin{figure*}[htp]
\centering
\includegraphics[height=0.55\columnwidth]{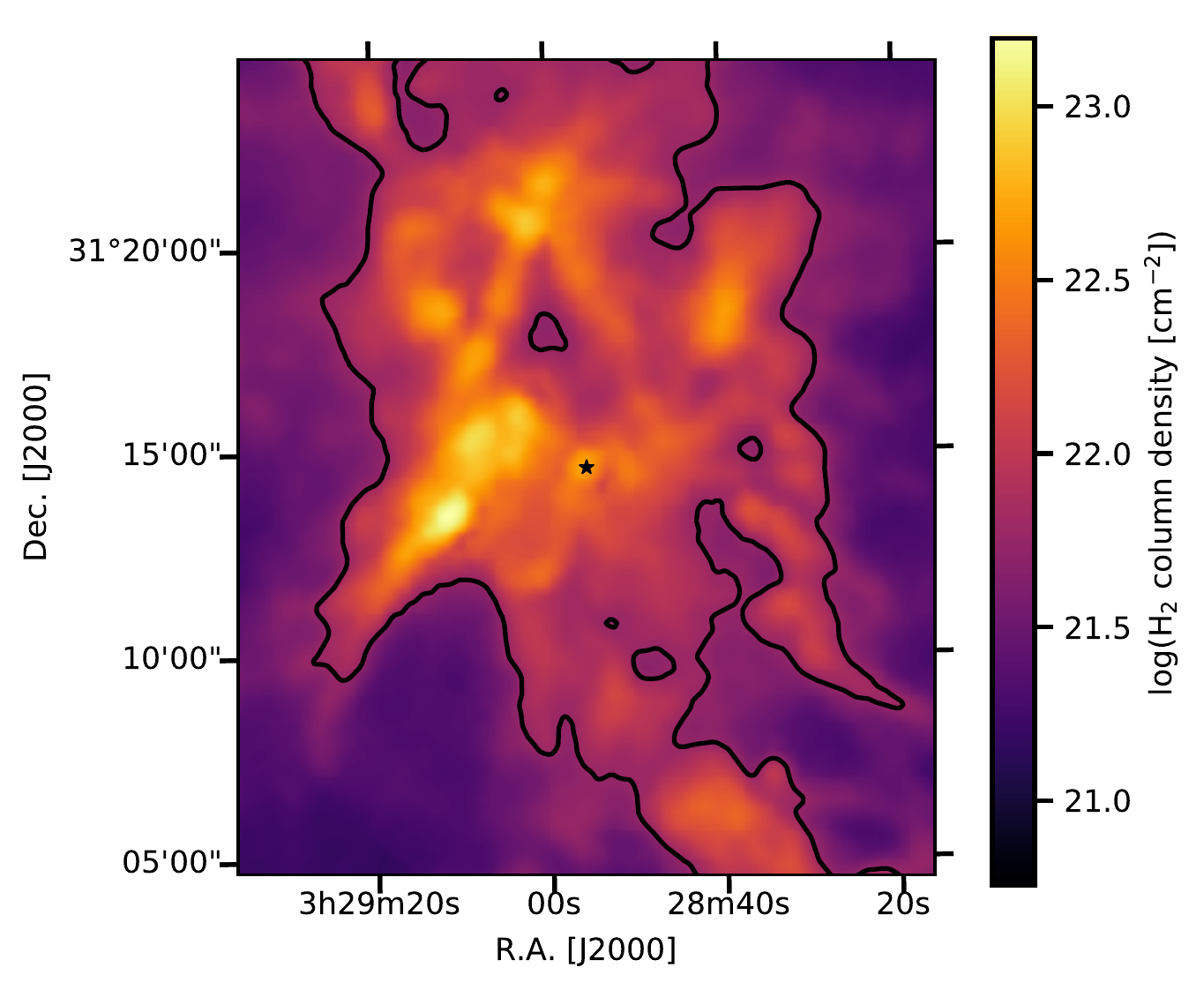} 
\includegraphics[height=0.55\columnwidth]{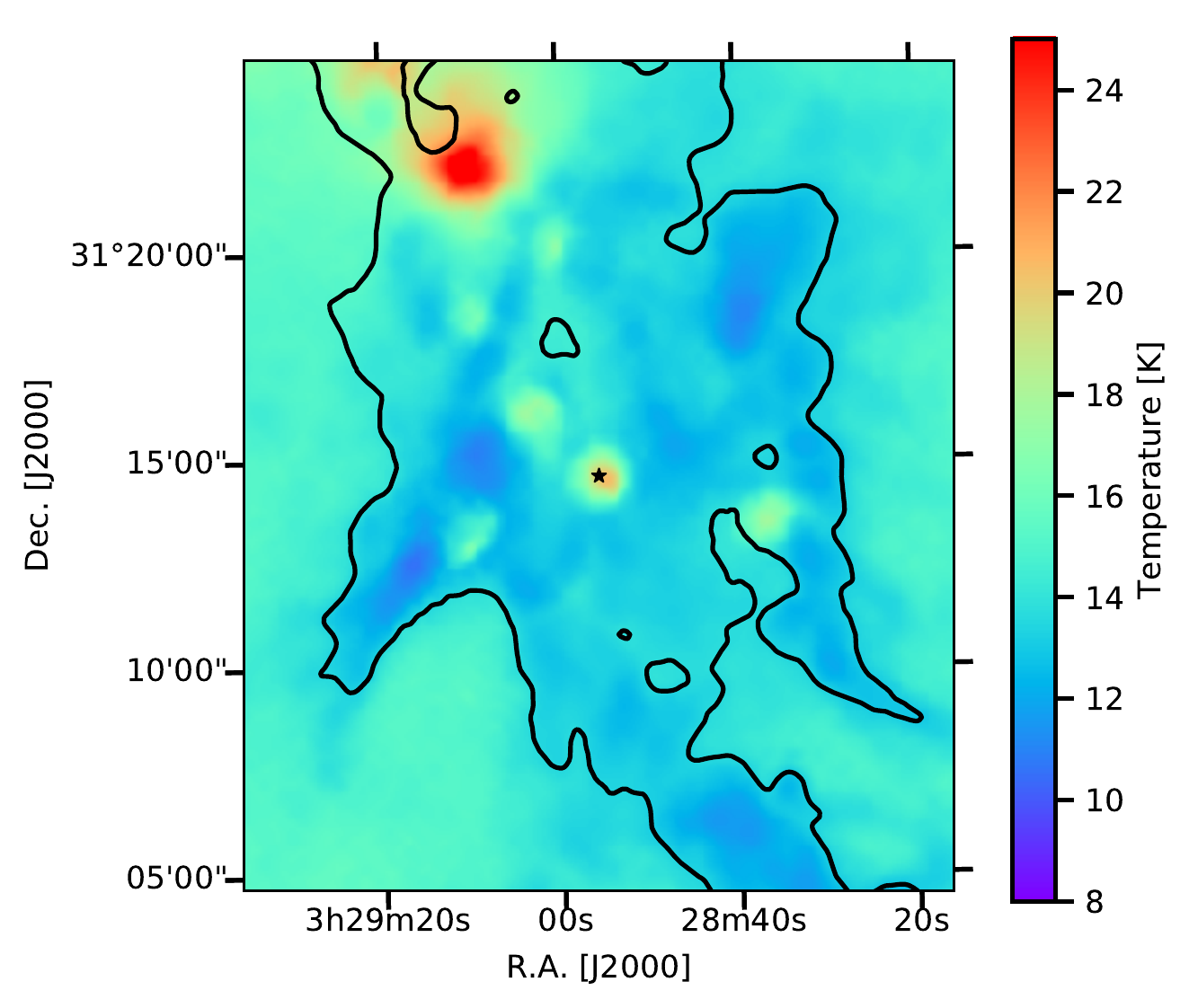} 
\includegraphics[width=0.66\columnwidth]{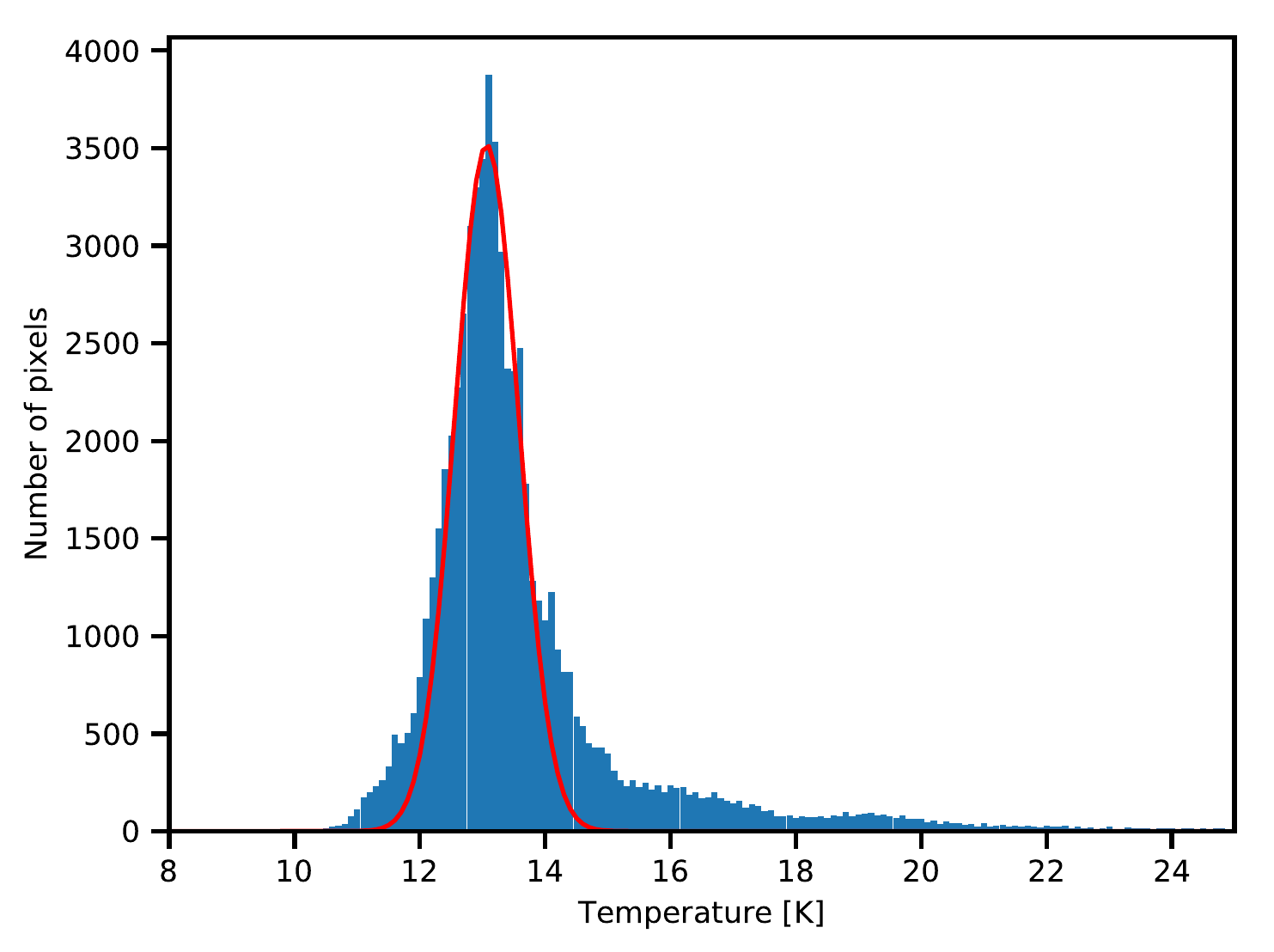} 
\caption{H$_2$ column density (left), dust temperature (center) maps, and dust temperature histogram of the map pixels with $A_{\rm V} > 3$ mag (right) of the NGC1333 star-forming region surrounding IRAS2A as observed with the {\it Herschel Space Observatory} by the Gould Belt survey \citep{Andre2010}. The black contour in the maps depicts the 3 mag level. The red curve on the histogram depicts the gaussian fit of the histogram around the histogram peak.}
\label{temp_cloud}
\end{figure*}

\section{Comparison with model predictions}

\begin{figure*}[htp]
\centering
\includegraphics[width=\textwidth]{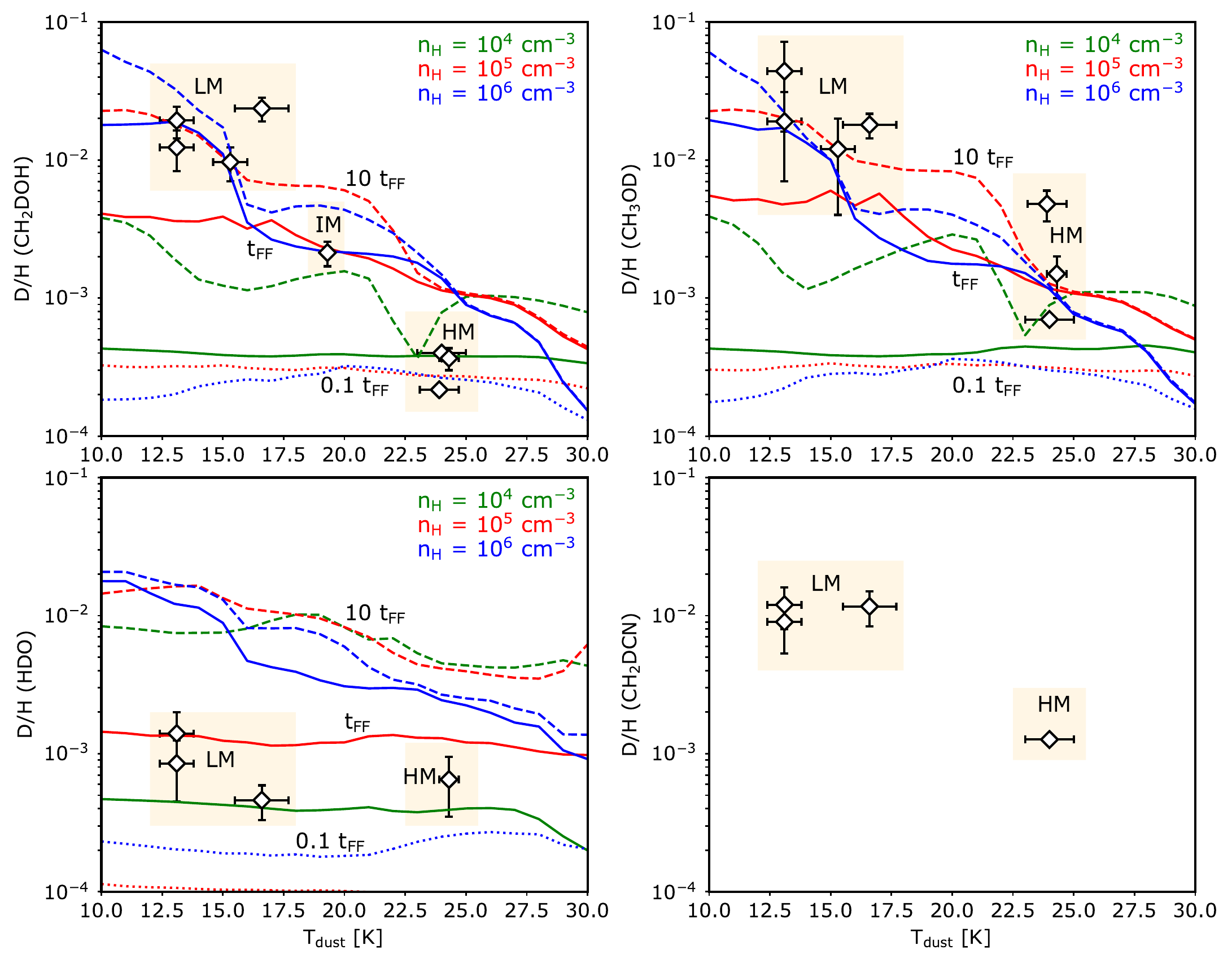} 
\caption{
{Statistical {(i.e., taking into account the statistical ratios for the functional groups with multiple H-atoms)} CH$_2$DOH (top left), CH$_3$OD (top right), HDO (bottom left), and CH$_2$DCN (bottom right) } deuterium fractionations observed towards the hot core of a sample of low-mass, intermediate-mass, and  high-mass protostars with sub-mm interferometers as a function of the dust temperature of the surrounding cloud measured with the {\it Herschel} or {\it Planck} telescopes (see values in Table \ref{table_Dmeth}). 
  Green, Red and Blue curves show the deuterations predicted by the \texttt{GRAINOBLE} astrochemical model in ices as
  function of the dust temperature assumed in the dark cloud simulation at $n_{\rm H} = 10^4$, $n_{\rm H} = 10^5$ and $10^6$ cm$^{-3}$, respectively for at $0.1 \times t_{\rm FF}$ (dotted lines), $t_{\rm FF}$ (solid), and $10 \times t_{\rm FF}$ (dashed) where $t_{\rm FF}$ is the free-fall time for the corresponding density.
  No predictions are given for the CH$_3$CN deuteration since this molecule is thought to be mostly produced in the vicinity of protostars either on ices or in the gas phase through warmer chemistry. "LM", "IM", and "HM" stand for Low-Mass, Intermediate-Mass, and High-Mass protostars, respectively.
  See text for more details.
  }
\label{deut_obs_mod}
\end{figure*}

\end{document}